\documentclass[aps,prl,reprint,amsmath,amssymb,superscriptaddress, 10pt]{revtex4-2}

\usepackage[usenames,dvipsnames]{xcolor}
\usepackage{amsmath,amssymb}
\usepackage{graphicx}
\usepackage{hyperref}
\usepackage{braket}
\usepackage[normalem]{ulem}
\usepackage{enumerate}
\usepackage{MnSymbol}

\newcommand{\be}{\begin{equation}}
\newcommand{\ee}{\end{equation}}
\def\bes{\begin{subequations}}
\def\esu{\end{subequations}}

\newcommand{\dr}{\text{dr}}
\newcommand{\bdr}{\textbf{dr}}
\newcommand{\eff}{\text{eff}}
\newcommand{\cc}{\text{c}}

\newcommand{\dd}{{\rm d}}

\newcommand{\M}{{\text{Z}}}
\newcommand{\R}{{\text{R}}}
\newcommand{\Pv}{{\mathcal{P}}}

\renewcommand{\selectlanguage}[1]{}
\renewcommand{\vec}[1]{\textbf{#1}}

\newcommand{\rhop}{{\varrho_\text{ep}}}
\newcommand{\rhoa}{{\varrho_\text{ea}}}

\newcommand{\Sp}{{\mathcal{E}}}

\newcommand{\abstractC}{The classical Landau--Lifshitz equation -- the simplest model of a ferromagnet -- provides an archetypal example for studying transport phenomena. In one-spatial dimension, integrability enables the classification of linear and nonlinear mode spectrum. An exact characterization of finite-temperature thermodynamics and transport has nonetheless remained elusive. We present an exact description of thermodynamic equilibrium states in terms of interacting modes. This is achieved by retrieving the classical Landau--Lifschitz model through the semiclassical limit of the integrable quantum spin$-S$ anisotropic Heisenberg chain at the level of the thermodynamic Bethe ansatz description. In the axial regime, the mode spectrum comprises solitons with unconventional statistics, whereas in the planar regime we find two additional types of modes of radiative and solitonic type. Our framework enables analytical study of unconventional transport properties: as an example we study the finite-temperature spin Drude weight, finding excellent agreement with Monte Carlo simulations.
}

\begin{document}

\newcommand{\titleinfo}{
 Landau-Lifschitz magnets: exact thermodynamics and transport}

\title{\titleinfo}

\author{Alvise Bastianello}
\affiliation{Technical University of Munich, TUM School of Natural Sciences, Physics Department, 85748 Garching, Germany}
\affiliation{Munich Center for Quantum Science and Technology (MCQST), Schellingstr. 4, 80799 M{\"u}nchen, Germany}
\author{\u{Z}iga Krajnik}
\affiliation{Department of Physics, New York University, 726 Broadway, New York, NY 10003, USA}
\author{Enej Ilievski}
\affiliation{Faculty for Mathematics and Physics, University of Ljubljana, Jadranska ulica 19, 1000 Ljubljana, Slovenia}

\begin{abstract}
\abstractC
\end{abstract}

\maketitle

\paragraph{\textbf{Introduction ---}} A quantitative understanding of macroscopic phenomena in interacting many-body systems is a central goal of theoretical and experimental physics. However, strong interactions make perturbative calculations unreliable, and to make progress, one has to identify appropriate collective degrees of freedom. An emblematic example of this paradigm are solitons, referring to stable particle-like coherent field excitations, found across various domains of physics including shallow water waves \cite{remoissenet2013waves}, gravity \cite{verdaguer2001gravitational}, cold-atom gases \cite{Eiermann2004,Morsch2006,Lannig2020}, magnets \cite{Ferrari2020,Raman2020} and others \cite{Tercas1,Tercas2,Tercas3}. Since the density of excited solitons is highly suppressed at low temperature, it has been suggested that thermodynamic quantities can be accessed by treating the system as a dilute gas, assuming solitons behave as well-separated quasiparticles \cite{Zakharov,Krumhansl1975} -- giving birth to the phenomenological \emph{soliton-gas approach} \cite{Currie_1980,Theodorakopoulos1984b,Kazuo1986,Sachdev1997,DamleSachdev2005,Kormos2016}.
While initially proposed only as an approximate technique for capturing physics at low temperature, it has been argued in subsequent works that in integrable models the soliton-gas description should provide accurate results even at finite temperature \cite{Currie_1980}, i.e. far from the dilute gas regime.

\begin{figure}[t!]
\centering
	\includegraphics[width=0.99\columnwidth]{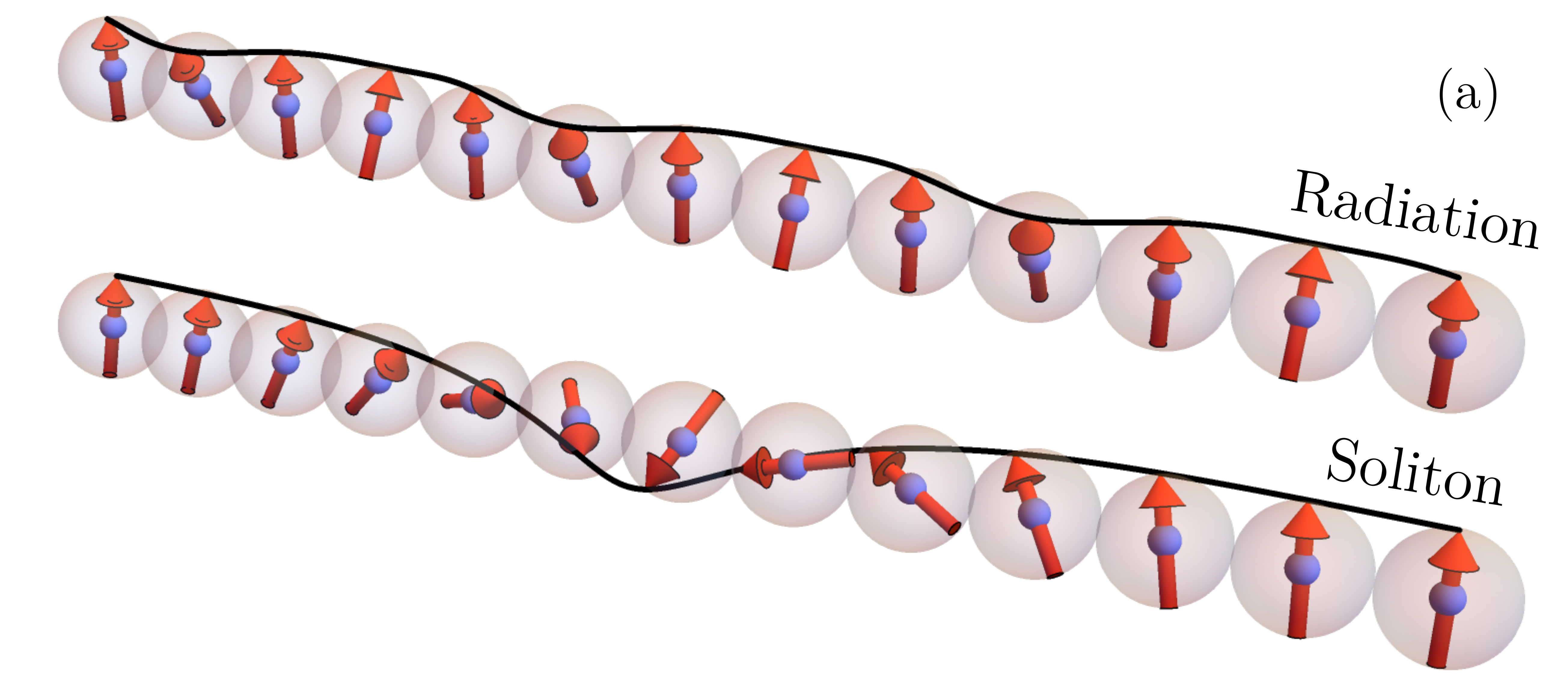}\\
\includegraphics[width=0.99\columnwidth]{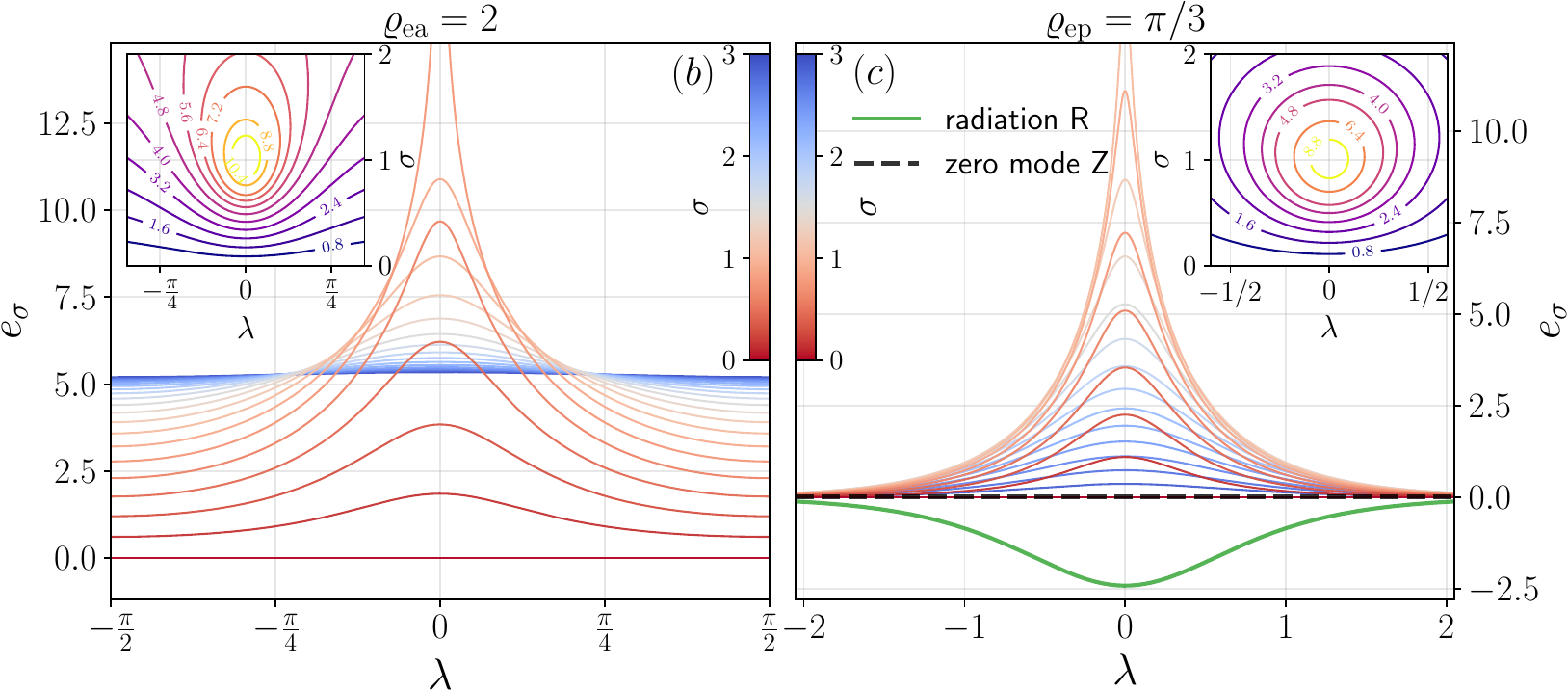}\caption{\textbf{Landau--Lifschitz excitations.---} $(a)$ The inverse scattering method decomposes a spin configuration in terms of delocalized radiation (akin to spin waves) and localized solitons as the fundamental excitations above the ferromagnetic vacuum. In the thermodynamic limit the mode structure depends on the regime, we focus on the lattice discretization \eqref{eq_latticeLLH}. $(b)$ The easy-axis regime supports a continuous family of solitons whose energy $e_\sigma(\lambda)$ is sharply peaked at $(\lambda, \sigma) = (0, 1)$ (see text for details). $(c)$ The easy-plane regime supports two additional types of modes: radiation R with negative energy and a zero mode Z with vanishing energy. Insets of Panels $(b)$, $(c)$ show energy contours in the two regimes.
}
	\label{fig_1}
\end{figure}

The inverse scattering method (ISM) \cite{ablowitz1981solitons,Faddeev_Takhtajan_1987,novikov1984} constitutes a general framework for studying classical integrable partial differential equations (PDEs). 
Within ISM, any field configuration that decays to the classical vacuum at spatial infinity can be uniquely decomposed in terms of delocalized radiative modes (often called phonons) and localized waves called solitons,
see Fig. \ref{fig_1}$(a)$ for a pictorial representation. A major downside of the ISM is that it cannot directly account for configurations with finite energy density that enter thermodynamic ensembles. 
Recent theoretical works \cite{EL2003,Zakharov2009,El_2021,suret2023soliton} in the domain of soliton gases either deal with special initial conditions or construct multi-soliton states with a prescribed set of parameters, without explicitly
relating them to thermodynamic state functions.
The conventional soliton-gas approach to thermodynamics stipulates that the partition sum can be evaluated exactly by performing a sum over the ISM excitation spectrum.
Interactions contribute solely by giving an effective length to quasiparticles, capturing the center of mass displacement after a scattering event \cite{Faddeev:1987ph}.
However, whether \emph{i)} the ISM modes truly provide an (over)complete set of degrees of freedom, and \emph{ii)} the nature of their statistical weights, both of which are pivotal for exact computation of thermodynamic properties, have remained open questions.

Most of the progress has so far been achieved in certain special cases, including the models that only feature radiative modes, such as the sinh-Gordon theory and defocusing non-linear Schr\"{o}dinger equation \cite{Bullough_1986}, or modes with one type of soliton mode with a single degree of freedom such as, for example, the Toda chain
\cite{Theodorakopoulos1995,Cao2019,Doyon2019toda,Bulchandani2019}, the KdV equation \cite{Bonnemain_2022}, and the Ablowitz–Ladik model \cite{Spohn2022,brollo2024particle}.
By contrast, generic models which involve both radiation and solitons, and may even comprise multiples species of nonlinear waves, are much more difficult to describe. Despite intensive efforts, the early attempts to obtain an exact soliton-gas description for the sine-Gordon model \cite{Timonen1986,Chen1986,Maki1985,Chung1989} (which has been achieved only recently in \cite{Koch2023}) and the Landau-Lifschitz equation \cite{Theodorakopoulos91,Theodorakopoulos95} have not come to fruition.
In fact, numerous inconsistencies and controversies \cite{Chung1990} hinted that such an approach might suffer from a fundemental conceptual flaw.

In the meantime, rapid advances in the manipulation of quantum matter \cite{Bloch2008,Bloch2012} have steered interest towards the study of quantum integrable systems \cite{Kinoshita2006,Haller2009,Langen2014,Schweigler2017,Wilson2020,Guan_2022}. The powerful tools of thermodynamic Bethe ansatz (TBA) \cite{takahashi2005thermodynamics} and generalized hydrodynamics (GHD) \cite{Doyon2016,Bertini2016,specialissueGHD} have delivered a myriad of exact results in various equilibrium and non-equilibrium settings, and led to important experimental confirmations \cite{Bouchoule_2022,Malvania2020,Moller2021,Cataldini2022,dubois2023}.
Crucially, the TBA formalism for quantum systems -- a quantum analogue of the soliton-gas approach -- is free of nuances that one typically encounters in classical integrable PDEs such as, for example, the determination of correct excitation statistics. This motivates the search for an alternative path to thermodynamics of classical integrable systems: taking the semiclassical limit directly at the level of TBA equations \cite{DeLuca2016,Bastianello2018,DelVecchio2020,Koch2022,Koch2023,bastianello2023sinegordon} avoids most of the pitfalls of the phenomenological soliton-gas approach.

In this Letter, we describe how to compute exact thermodynamic properties of the classical Landau-Lifschitz (LL) model -- a prototypical model for a ferromagnet --
and derive hydrodynamic equations governing the evolution on the ballistic scale.
This is achieved by regarding the LL model as the large-spin limit of the integrable quantum spin-$S$ chains \cite{KR87I,KR87II,KRS90,Frahm90,Bytsko03}. While semiclassical limits of integrable quantum magnets have in some capacity been addressed previously in the literature \cite{Sklyanin88,Krajnik2021,Miao21,Haldane1982,DeNardis2020}, we here for the first time manage to determine the complete set of modes, alongside their associated statistical weights, relevant for capturing exact thermodynamic properties.
Specifically, our description of thermal Gibbs states (or any generalized Gibbs Ensembles \cite{Rigol2007}) is embedded in the standard TBA framework, thus providing the long sought soliton-gas picture.
Most remarkably, unlike radiation and solitons which are conventionally associated with Rayleigh-Jeans and Maxwell-Boltzmann statistics, respectively, the modes obey unorthodox statistics.

Our results provide a direct access to many applications in physics. Here we wish to particularly highlight the recent discovery of spin superdiffusion in integrable quantum~\cite{Marko11,Ljubotina19,Ilievski18,GV19,DeNardis2020,superuniversality,superdiffusion_review} and classical spin chains \cite{Das2019,Krajnik2020}, anomalous spin-current statistics \cite{Krajnik2021,SarangFCS,Krajnik2022,krajnik2023dynamical}, 
and anomalous type of transport in nonintegrable classical chains~\cite{DasDamleDharHuseKulkarniMendlSpohn2019,mcroberts2023,roy2023}, which have also attracted significant interest in experimental communities \cite{Jepsen2020,Tennant2021,Wei2022,Rosenberg2024}.
Our work offers a theoretical and computational framework for a detailed investigation of nonequilibrium phenomena, and anomalous transport in particular for which, in spite of tremendous advancements, a complete understanding is still lacking.

\paragraph{\textbf{The model---}} 
The classical field theory of a one-dimensional classical Landau-Lifschitz magnet is governed by the Hamiltonian
\be\label{eq_LLcH}
H_{\rm LL}=\frac{1}{2}\int_{\mathbb{R}} \dd x \,\big[ (\partial_{x}\textbf{S})^{2}+\varDelta(1-(S^{3})^{2})\big],
\ee
where the spin field $\textbf{S}(x)=(S^{1}(x),S^{2}(x),S^{3}(x))$, normalized to unity $\textbf{S}\cdot \textbf{S}=1$, obeys the Lie--Poisson brackets $\{S^a(x),S^b(y)\}=\Sp\epsilon_{abc}S^{c}(x)\delta(x-y)$ and $\Delta \in \mathbb{R}$ is the anisotropy. Here $\Sp$ is a free parameter used to set the timescale in the equation of motion $\dd \vec{S}/\dd t = \{\vec{S},H\}$. We pick $\Sp=2$ which will be justified later on. The third component of total magnetization, $M^{3}\equiv \int \dd x\, S^{3}(x)$, is conserved in time. In the easy-axis regime $\varDelta>0$, the lowest energy configuration is the ferromagnetic vacuum with constant magnetization $S^3(x)=\pm 1$, while in the easy-plane $\varDelta<0$ regime all spins align in the orthogonal plane with $S^3(x)=0$.

\begin{figure}[t!]
\centering
 \includegraphics[width=0.99\columnwidth]{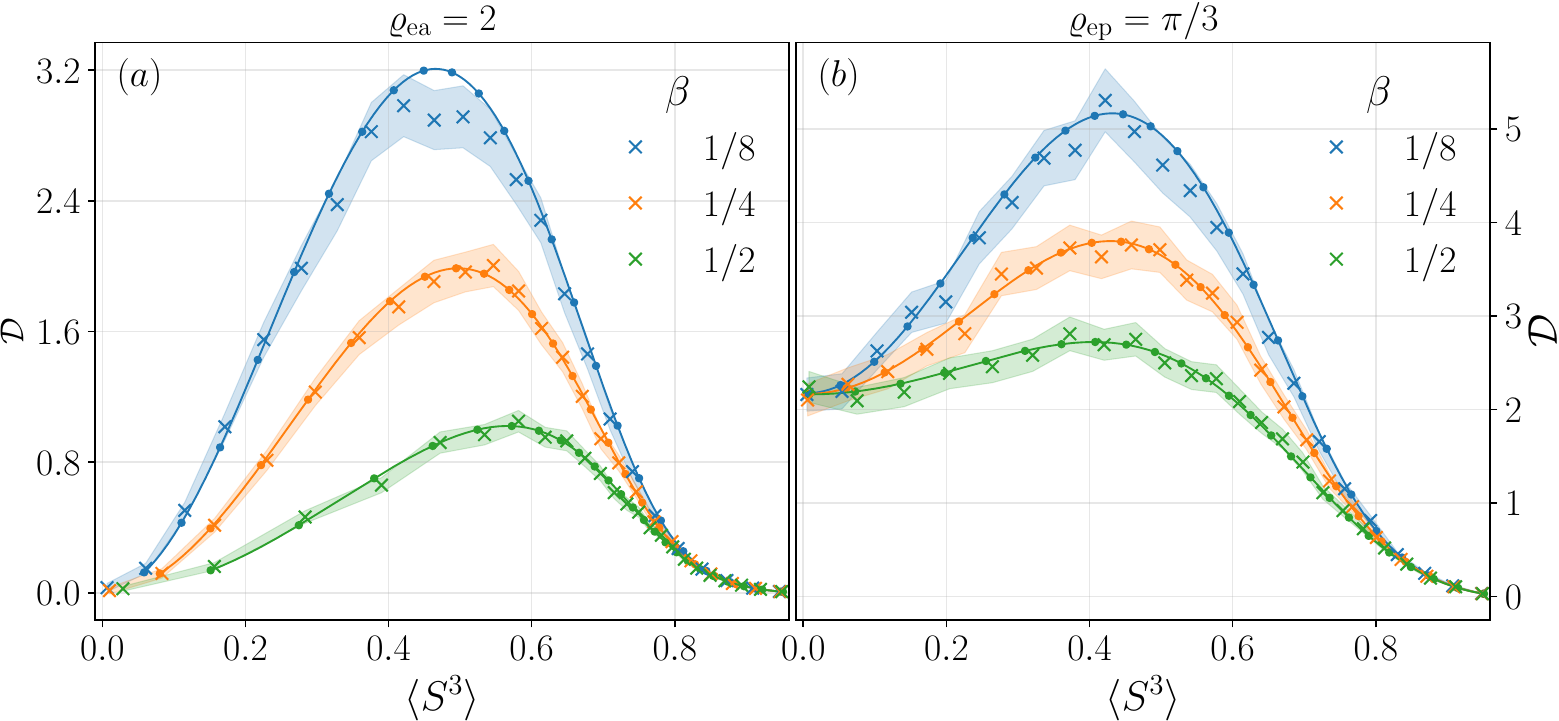}
 \caption{
 \textbf{Spin Drude weight.---}
 We consider the lattice Landau-Lifschitz \eqref{eq_latticeLLH} and compare the result of the TBA equations (circles, solid lines as a guide to the eye) with Monte Carlo data (crosses, with shaded regions showing two standard deviation confidence intervals). Panel $(a)$: easy-axis regime with $\varrho_{\rm ea} = 2$. Panel $(b)$: easy-plane regime with $\varrho_{\rm ep} = \pi/3$.
 Simulation parameters: system size $L=2\times 10^3$, number of samples $N = 2\times 10^3$, time-step $\tau = 0.03$, see Ref.~\cite{Krajnik2021} for details.
 }  
	\label{fig_2}
\end{figure}

Hereafter, we consider grand-canonical equilibrium ensembles with partition functions $Z(\beta,\mu) \equiv \int \mathcal{D}[\vec{S}]\,e^{-\beta H+\mu (M^{3}-M^3_\text{vac})}$
(normalized to $Z(0,0)=1$) where $M^3_\text{vac}$ denotes the average of $M^3$ in the ferromagnetic vacuum.
To establish the validity of our analytical description and to showcase its predictive power, we perform several independent checks. By imposing periodic boundary conditions for a system of length $L$, we first compute the free energy density $f=-\lim_{L\to \infty}L^{-1}\log Z(\beta,\mu)$, encoding the charge averages (and their static correlation functions) such as, for example, the average magnetization $\langle S^3\rangle$. As a paradigmatic probe of transport, we consider dynamical correlation functions of the spin-current and compute the associated Drude weight
$\mathcal{D}=\lim_{t\to\infty}\int \dd x\, \langle j(x,t) j(0,0)\rangle_{\text{c}}$ where the spin current density $j(x,t)$ is defined via $\partial_{t} S^{3}(x,t)+\partial_{x}j(x,t)=0$.
Recall that integrable models in general feature finite non-zero Drude weights, attributed to ballistic propagation of quasiparticle excitations \cite{De_Nardis_2022}. Both probes, namely average magnetization and spin Drude weight, provide nontrivial checks for the completeness of the derived classical TBA equations. 
Specifically, in the infinite-temperature limit we analytically retrieve the magnetization curve $\langle S^3\rangle=-1/\mu+\coth\mu$ from our TBA \cite{suppmat}. 
Additionally, we find excellent agreement with numerical simulations at finite temperatures, see Fig.~\ref{fig_2} and SM \cite{suppmat} for the Drude weight and magnetization respectively.

Conveniently, the LL model \eqref{eq_LLcH} admits an integrable lattice discretization \cite{Faddeev_Takhtajan_1987}, the lattice Landau-Lifschitz (LLL) model, $H_{\text{LLL}} = -2 \sum_\ell \log \varPhi_{\ell, \ell+1}$, with 
\begin{multline} \label{eq_latticeLLH}
\varPhi_{\ell,\ell+1}=
\Upsilon(S^3_\ell)\Upsilon(S^3_{\ell+1})
\Big(S^1_\ell S^1_{\ell+1} +S^2_\ell S^2_{\ell+1}\Big)+\\
U(1)U\big(\tfrac{1}{2}(S^{3}_{\ell}+S^{3}_{\ell+1})\big)
-U\big(\tfrac{1}{2}(S^{3}_{\ell}-S^{3}_{\ell+1})\big),
\end{multline}
with auxiliary functions $U(y)\equiv \cosh{(\varrho_{\rm ea}y)}$ and $\Upsilon(y) \equiv \sqrt{\big(U(1) -U(y)\big)/\big(1-y^2\big)}$, and the easy-axis anisotropy parameter $\varrho_{\rm ea} \in \mathbb{R}_{+}$. The easy-plane regime is reached by analytic continuation $\varrho_{\rm ea} \to i\,\varrho_{\rm ep}$, $\varrho_{\rm ep}\in [-\pi,\pi]$. 
The continuum limit, yielding Eq.~\eqref{eq_LLcH}, is recovered at large wavelengths by introducing the lattice spacing $a$, expanding $\vec{S}_{\ell\pm 1}=\vec{S}(x)\pm a\,\partial_{x}\vec{S}(x)+\mathcal{O}(a^{2})$, rescaling the interaction as $\varrho_{\rm ea} = a\sqrt{\varDelta}$, and letting $a\to 0$.
Since the field theory is accessible as a limit of the lattice Hamiltonian \eqref{eq_latticeLLH}, we subsequently focus our considerations on the lattice model, which is also more convenient for performing numerical simulations.

Since the final compact results can be discussed without dwelling on the details of taking the semiclassical limit (which follows previous, albeit simpler, derivations \cite{Koch2022,Koch2023}), we leave unessential details to the Supplementary Material (SM) \cite{suppmat} and discuss their general aspects and results.

\paragraph{\textbf{Thermodynamics of integrable models.---}}
We first briefly introduce the setting of TBA and define thermodynamic state functions (for more details, see SM \cite{suppmat} or the literature \cite{takahashi2005thermodynamics}).
Individual modes (excitations) are assigned a type (or specie) index ``$I$". The associated (bare) energy and momentum are denoted by $e_I(\lambda)$ and $p_I(\lambda)$ respectively, conveniently parametrized in terms of the rapidity variable $\lambda$. In spin chains, a type-$I$ excitation carries $m_I$ quanta of magnetization (relative to the ferromagnetic vacuum). In classical magnets, the two degrees of freedom of classical solitons (e.g. magnetization and momenta) take \emph{continuous} values. For compactness, we introduce an implicit notation for the scalar product $a_{I}\circ b_{I}$ and convolution $a_{I}\star b_{I}$, evaluated over the rapidity domain for any two quantities $a_{I}$ and $b_{I}$ ascribed to specie $I$, while simultaneously adopting the summation convention in the case of repeated index $I$ (i.e. integration in the case of $I$ having a continuous range, see SM \cite{suppmat} for details).

Gibbs ensembles, or more generally generalized Gibbs ensembles \cite{Rigol2007}, are uniquely identified by the rapidity densitie $\rho_I(\lambda)$.
The total densities of available states are given by $\rho_I^t=\tfrac{\kappa_{I}}{2\pi}(\partial_\lambda p_I)^\dr$ with $\kappa_I=\text{sign}[\partial_\lambda p_I]$, and represents the effective available phase space for each mode. Owing to
interactions, quantities $g_I$ associated to mode $I$ gets renormalized. This effect is known as dressing, $g_I\mapsto g_I^\dr$, and amounts to solving a \emph{linear} (Fredholm) integral equation of the form
$g^{\dr}_{I}+T_{I,I'}\star (\kappa\vartheta g^{\dr})_{I'}=g_I$
where $T_{I,I'}$ encode the effect of interaction among the species of type $I$ and $I'$ (related to the time delay \cite{Doyon2018,Doyon2023,Faddeev_Takhtajan_1987} induced by scattering), while the filling fractions are given by the ratios $\vartheta_I \equiv \rho_I/\rho_{I}^t$. 

The mode densities $\rho_{I}$ of the Gibbs state can be inferred by minimizing the free energy $F=\beta \langle H\rangle-\mu\big(\langle M^3\rangle -M^{3}_{\rm vac}\big)-\mathcal{S}$.
Here $\mathcal{S}$ denotes the thermodynamic entropy, obtained by summing over all modes with appropriate statistical weights $s_I= s_I(\vartheta_I)$. Writing the entropy density as $s \equiv \lim_{L\to \infty}(\mathcal{S}/L)=\rho^t_I \circ s_I$ yields the following spectral resolution of the free-energy density \cite{suppmat}
$
f=\lim_{L\to \infty}(F/L)=\tfrac{1}{2\pi}\kappa_{I}\partial_\lambda p_I\circ \mathcal{F}_I
$, with $ \mathcal{F}_I(\vartheta_I) \equiv \vartheta_I s'_I(\vartheta_I)-s_I(\vartheta_I)$. On Gibbs ensembles, the occupations $\vartheta_{I}$ satisfy the following nonlinear integral equations
\be\label{eq_TBA}
s'_I(\vartheta_I)=\beta \, e_I+\mu \, m_I-T_{I,I'}\star \kappa_{I'}\mathcal{F}_{I'}(\vartheta_{I'})\,,
\ee
where $s'_I(\vartheta)=\dd s_I(\vartheta)/\dd \vartheta$.
Once $\rho_{I}$ have been determined, analytically or numerically, the average charge densities can be simply computed by multiplying them with bare charges and summing over the entire spectrum.

For instance, the magnetization density is given by $\langle S^3\rangle=L^{-1}M_{\rm vac}+m_I\circ\rho_I$. On the other hand, the spin Drude weight assumes the following mode resolution \cite{Doyon2017,Doyon2018cor,De_Nardis_2022,suppmat}
\be\label{eq_ghdD}
\mathcal{D}=\rho_I w_I \circ (m_{I}^{\dr}\,v^{\rm eff}_{I})^2,
\ee
where $w_I(\vartheta_I)\equiv -1/[\vartheta_I s''_I(\vartheta_I)]$ accounts for the mode statistics and the effective velocity is $v_I^\eff \equiv (\partial_\lambda e_I)^\dr/(\partial_\lambda p_I)^\dr$. The latter identification is sensitive to the choice of time scale in the equation of motion, and our choice $\Sp=2$ ensures its validity, see SM \cite{suppmat}.

The functional form of statistical factors $s_I$ discerns the nature of quasiparticle modes: typically in quantum models one finds the Fermi-Dirac statistics $s_{\text{FD}}(\vartheta)=-\vartheta\log\vartheta-(1-\vartheta)\log(1-\vartheta)$ \cite{takahashi2005thermodynamics}. 
By contrast, in classical systems one usually encounters radiation $s_{\text{Rad}}(\vartheta)=\log \vartheta$ \cite{Bastianello2018,DelVecchio2020} with Rayleigh-Jeans statistics, or solitons $s_{\text{Sol}}(\vartheta)=\vartheta(1-\log \vartheta)$ \cite{Cao2019,Doyon2019toda,Bonnemain_2022} with the associated Boltzmann weight. 
As detailed out in the remainder of the paper, the LLL model evades this simple description: we find solitons with unorthodox (renormalized) statistical weights alongside (depending on the regime of anisotropy) radiation and exceptional non-dynamical (zero-energy) solitons.

\paragraph{\textbf{Easy-axis regime.---}} We first inspect the easy-axis regime of the Hamiltonian \eqref{eq_latticeLLH}. More details, including the continuum limit \eqref{eq_LLcH}, can be found in the SM \cite{suppmat}.
In this regime, rapidities occupy a compact domain $\lambda\in [-\pi/2,\pi/2]$, and the mode spectrum comprises solely solitons with bare energy $e_\sigma(\lambda)$, labeled by $\lambda$ and a \emph{continuous} internal variable $\sigma\in \mathbb{R}_{+}$ associated with magnetization $m_\sigma=2\sigma$, and with positive parity $\kappa_\sigma=1$.
A typical dispersion law is shown in Fig. \ref{fig_1}$(b)$. Sectors with different magnetization signs are disconnected, and the ferromagnetic vacuum reference state must be chosen accordingly \cite{Piroli2017,suppmat}. The power-law singularity at $(\sigma,\lambda)=(1,0)$ is a consequence of the logarithmic interaction in Eq.\eqref{eq_latticeLLH}. 
Solitons acquire \emph{renormalized} Boltzmann weights
\be
s_\sigma(\vartheta_\sigma)=s_{\text{Sol}}(\vartheta_\sigma)-\sigma^{-2}-\vartheta_\sigma \log\sigma^2.
\ee
Although the form of $s_\sigma$ affects the filling fraction \eqref{eq_TBA}, we note that the additional terms (constant or linear in $\vartheta_\sigma$) do not affect the function $w_I$ appearing in the Drude weight \eqref{eq_ghdD}. The kinematic data retrieved by taking the semiclassical limit is compatible with expressions derived using the ISM (see SM \cite{suppmat} for explicit expressions).
In Fig.~\ref{fig_2}$(a)$ we show the Drude weight obtained by numerically solving the classical TBA equations and independently by performing Monte-Carlo simulations \cite{Metropolis2004,Hastings1970}, see also SM \cite{suppmat}. 
The divergence of the bare energy $e_\sigma(\lambda)$ at $(\sigma,\lambda)=(1,0)$,  see Fig.~\ref{fig_1}$(b)$, causes a singularity in the effective velocity $v^\eff$. 
The singularity is balanced by a zero of the filling function $\vartheta_\sigma(\lambda)\propto e^{-\beta e_{\sigma}(\lambda)}\propto ((\sigma-1)^2\varrho_{\rm ea}^{2}+4\lambda^2)^{2\beta}$, rendering the Drude weight finite.
However, this `damping' mechanism diminishes with decreasing $\beta$, resulting in a logarithmic divergence of the Drude weight at high temperatures, $\mathcal{D}\propto \log\beta^{-1}$ \cite{suppmat}.

\paragraph{\textbf{Easy-plane regime.---}}

Remarkably, in the easy-plane regime there appear \emph{three} distinct types of quasiparticles (above the ferromagnetic vacuum): a continuum of magnetic solitons with renormalized statistics (analogous to those in the easy-axis regime) and $\sigma \in [0,\pi/\varrho_{\rm ep}]$,
a single radiative mode $\R$ with $m_\R=2$ and renormalized entropy $s_\R(\vartheta_\R)=s_{\text{Rad}}(\vartheta_\R)+1+\log(\varrho_{\rm ep}/\pi)$ and, finally, a special type of soliton mode $\M$ with no bare energy nor momentum, $e_\M=\partial_\lambda p_\M=0$, characterized by finite magnetization $m_\M=2\pi/\varrho_{\rm ep}$ and renormalized entropy weight $s_\M(\vartheta_\M)=s_{\text{Sol}}(\vartheta_\M)+\vartheta_\M\log(\varrho_{\rm ep}/\pi)$.
Solitons have positive parity $\kappa_\sigma=1$ and $\kappa_\M=1$, whereas the radiative mode has $\kappa_\R=-1$.
In the planar regime, rapidities span the whole real line, $\lambda\in \mathbb{R}$, while the explicit expressions for dispersion laws and scattering kernels are reported in SM \cite{suppmat}.
The dispersion laws of these modes are represented in Fig. \ref{fig_1}$(c)$, featuring the same type of singularity as in the easy-axis regime, responsible for a logarithmic divergence of the Drude weight in the high-temperature limit.
In Fig.~\ref{fig_2}$(b)$ we compare the Drude weight \eqref{eq_ghdD} with our Monte Carlo numerical data.

\paragraph{\textbf{The spin-wave limit.---}}
Small fluctuations of the ferromagnetic vacuum can be expanded in terms of non-interacting delocalized spin waves, akin to phonons. Such spin waves are not explicitly present in our thermodynamic description and hence should somehow emerge out of the existing set of excitations.
The natural expectation is that very extended solitons should be practically indistinguishable from delocalized modes.
Indeed, a direct analytic calculation \cite{suppmat} demonstrates that in the easy-axis regime spin waves are retrieved as a cumulative effect of shallow solitons by summing over all $\sigma$ at fixed rapidity. 
The same applies for the positive-energy radiation branch in the easy-plane regime. By contrast, the radiative branch with negative energy must be included in the TBA as an independent mode in the easy-plane regime.
Note that $\M$ modes carry finite magnetization and are effectively eliminated at low densities.

\paragraph{\textbf{Discussion---}} By performing the
semiclassical limit of the integrable quantum spin-$S$ chain within the framework of the thermodynamic Bethe ansatz, we obtained the exact thermodynamics and hydrodynamics of the classical Landau--Lifshitz model. The obtained integral equations encoding thermodynamics of the model are consistent with a soliton-gas description, but they display several unexpected features. 
Firstly, in the easy-axis regime we find a spectrum of solitons with modified Boltzmann statistics, whereas radiative modes (spin waves) do not appear as independent modes; instead they can be seen as a condensate of wide solitons with small amplitude.
The easy-plane regime is even more peculiar: apart from solitons, the spectrum of modes includes localized zero-energy modes and only the negative-energy branch of spin waves.

This Letter brings to the front a number of important questions.
The most pressing one concerns the general validity of the soliton-gas approach to thermodynamics, which crucially relies on the stipulated form of statistical weights associated with different types of excitations. Our findings indicate that the anticipated Maxwell-Boltzmann or Rayleigh-Jeans statistics are not always appropriate and thus it is unsafe to apriori assume them. Presently we are only able to corroborate this claim through a systematic semiclassical analysis of quantum spectra, while an independent purely classical justification using the tools of inverse scattering \cite{Faddeev_Takhtajan_1987,AblowitzBook} and finite-gap integration \cite{Kamchatnov92,Kamchatnov2000,Kazakov2004} is still lacking.

Our results are expected to facilitate further progress in understanding anomalous transport phenomena in integrable magnets
and help elucidating the elusive phenomenon of spin superdiffusion at the isotropic point and its
intimate connection with the Kadar--Parisi--Zhang to the universality class: while the dynamical exponent and scaling function are by now firmly established, \cite{ZunkovicProsen13,Ilievski18,Ljubotina19,Das2019,Krajnik2020,Evers2020,Moore2020,superuniversality}, the
spin-current fluctuations  \cite{Krajnik2022,krajnik2023dynamical,Jacopo_KPZ_NLFHD} reveal a discernibly distinct behavior. Computing the full counting statistics of charge transport \cite{BMFT,SpaceTimeDuality,Doyon2023a} in the LL magnets using the derived classical TBA equations might lead to important new insights. There are are many other interesting questions to address such as e.g.~obtaining nonabelian hydrodynamics that governs the evolution of gauge modes associated to polarization direction of the ferromagnetic vacuum at the isotropic point \cite{Bulchandani2020}, or the study of thermalization in the presence of integrability breaking \cite{Bastianello_2021} with multiple quasiparticle species. Addressing these questions requires both a fully-fledged analytical toolbox and extensive numerical benchmarks. The results of our work thus make the classical Landau--Lifschitz model an ideal playground for realizing this program.

\paragraph{\textbf{Data and code availability---}} Raw data and working codes are available on Zenodo \cite{Zenodo}. 
\paragraph{\textbf{Acknowledgments---}}
We thank Jacopo De Nardis, Oleksandr Gamayun, Tomaž Prosen, Herbert Spohn and Takato Yoshimura for useful discussions and comments on the manuscript.
AB acknowledges support from the Deutsche Forschungsgemeinschaft (DFG, German Research Foundation) under Germany’s Excellence Strategy–EXC–2111–390814868.  \v{Z}K is supported by the Simons Foundation via a Simons Junior Fellowship grant 1141511.
EI is supported by the Slovenian Research Agency (ARIS) grants P1-0402 and N1-0243.

\bibliography{biblio}

\newpage

\onecolumngrid

\setcounter{equation}{0}  
\setcounter{figure}{0}
\setcounter{page}{1}
\setcounter{section}{0}    
\renewcommand\thesection{\arabic{section}}    
\renewcommand\thesubsection{\arabic{subsection}}    
\renewcommand{\thetable}{S\arabic{table}}
\renewcommand{\theequation}{S\arabic{equation}}
\renewcommand{\thefigure}{S\arabic{figure}}
\setcounter{secnumdepth}{2}  

\begin{center}
{\Large \textbf{Supplementary Material}}\\ \ \\
{\large \textbf{\titleinfo}}
\ \\ \ \\
Alvise Bastianello, \u{Z}iga Krajnik, Enej Ilievski
\end{center}
\bigskip
\bigskip

This Supplementary Material covers the technical aspects, including the derivations, presented in the Letter. In particular:
\begin{enumerate}
\item In Section \ref{sec_quantum} we review the TBA description of the integrable quantum spin-$S$ chains.
\item In Section \ref{sec_SW} we provide the spin-wave limit analysis of the classical spin chain. This calculation is crucial to fix the proper normalization of the Lie-Poisson brackets that allows to identify the spin current with the usual GHD identities for the Drude weight.
\item In Section \ref{sec_semiclassical} we discuss in details the semiclassical limit of the TBA of the quantum spin chain.
\item In Section \ref{sec_numerics} we describe the numerical methods used in solving the TBA equations, and performing the microscopic simulations.
\end{enumerate}

\bigskip
\textbf{Notation for scalar products and convolutions in the rapidity space. --- } In the main text, we introduced the short-hand notation for the scalar product $f_I\circ g_I\equiv \sum_{I'} \int \dd\lambda f_{I}(\lambda) g_{I}(\lambda)$ for arbitrary functions $f_I$ and $g_I$, where the summation over the modes $I$ is replaced by an integral in the case of a continuous spectrum. Instead, the convolution $T_{I I'}\star f_{I'}$ is defined as $[T_{I I'}\star f_{I'}](\lambda)\equiv \sum_{I'} \int \dd \lambda'\, T_{I I'}(\lambda-\lambda') f_{I'}(\lambda')$.

\section{The integrable quantum spin-$S$ chains}
\label{sec_quantum}

In this Section, we provide a short summary of the excitation spectra and thermodynamics of the integrable quantum spin-$S$ chains. We shall use
these result as the starting point of our analysis In Section \ref{sec_semiclassical} where we systematically take the semiclassical limit of the quantum model and deduce the equations governing the classical thermodynamics of the Landau--Lifshitz spin chain.

The higher-spin analogue of the Heisenberg spin-$1/2$ chain can be obtained by employing the fusion technique \cite{KR87I,KR87II}. An infinite tower of local Hamiltonians can then be obtained in the usual fashion by taking the logarithmic derivatives of the fused commuting transfer matrix evaluated at a particular point. We shall not review the procedure here, since the spin-$S$ Hamiltonians are given by rather cumbersome expressions, besides being inessential for the implementation of our program. Instead, the reader is referred to Refs. \cite{Frahm90,Bytsko03} and references therin.

\medskip

\textbf{Thermodynamic Bethe Ansatz. ---}
The exact thermodynamics of integrable models is describable within the Thermodynamic Bethe Ansatz (TBA) framework \cite{takahashi2005thermodynamics}. The general structure is as follows: the elementary excitations are magnons with bare momentum $p$. We use a convenient parametrization in terms of the rapidity variable $\lambda\in(-\Lambda,\Lambda)$. The cutoffs in the rapidity space $\Lambda$ and the number of strings are model-dependent. Due to attractive interaction, magnons can bind into coherent bound states called Bethe strings. The internal quantum number of a string, $j\in \{1,2,...,N\}$, is the binding number, i.e. the number of magnetization quanta -- the $U(1)$ charge of a string.
Each magnon excitation carries finite amount of (bare) charges. The total charge of an eigenstate is obtained by adding up the charges of individual magnons or strings.
The bare energy, momentum and $U(1)$ charge (magnetization) are denoted by $e_j(\lambda)$, $p_j(\lambda)$ and $m_{j}$, respectively.
In the thermodynamic limit, one describes macrostates with the so-called root density, which is interpreted as the phase-space density of these excitations and is denoted as $\rho_j(\lambda)$.
In the thermodynamic limit, the total energy density, magnetization and more general conserved quantities can be expressed weighting the summation over the bare charges with the excitations' root density. For example, in an homogeneous system of size $L$ one has
\be
\lim_{L\to\infty}\frac{1}{L}\langle H\rangle=\sum_{j=1}^N \int_{-\Lambda}^{\Lambda}\dd\lambda \, e_j(\lambda)\rho_j(\lambda)\, ,\hspace{2pc}\lim_{L\to\infty}\frac{1}{L}\langle S^z\rangle=1-\sum_{j=1}^N \int \dd\lambda \, m_j \rho_j(\lambda)
\ee

Due to interactions, the allowed phase space for each excitation is renormalized by the presence of other quasiparticles. Hence, one introduces the total root density $\rho^t_j(\lambda)$ and the filling fraction $\vartheta_j(\lambda)=\rho_j(\lambda)/\rho_j^t(\lambda)$. The root density and total root densities are not independent, but are connected through the integral equation
\be
\kappa_j\rho^t_j(\lambda)=\frac{\partial_\lambda p_j(\lambda)}{2\pi}-\sum_{j'} \int_{-\Lambda}^{\Lambda} \dd\lambda'\, T_{j,j'}(\lambda-\lambda')\rho_{j'}(\lambda')\,, 
\ee
where $\kappa_j=\text{sign}(\partial_\lambda p_j)$ is the parity of the string and it is equal to the sign of $\partial_\lambda p_j(\lambda)$. $T_{j,j'}(\lambda)$ is the scattering kernel that accounts for interactions. Indeed, in the zero density limit $\rho_j\to 0$, and thus in the absence of scattering, the total root density becomes the bare one $|\partial_\lambda p_j|/(2\pi)$.

To obtain the root density that describes a thermal state, one needs to minimize the proper free energy. To this end, one introduces the Yang-Yang entropy
\be
\mathcal{S}=L\sum_{j=1}^N\int_{-\Lambda}^{\Lambda}\dd\lambda \rho_j^t(\lambda) s(\vartheta_j(\lambda))\, ,
\ee
where $s(x)$ is the entropy density. In quantum integrable models, the excitations have fermionic statistics and thus $s(x)=-x\log x-(1-x)\log(1-x)$.
Thermodynamics is determined upon maximizing the entropy constrained to energy and magnetization conservation, reaching the equations \cite{takahashi2005thermodynamics}
\be
\varepsilon_j(\lambda)=\beta e_j(\lambda)+\mu m_j+\sum_{j'=1}^N\int_{-\Lambda}^{\Lambda}\dd\lambda'  T_{j,j'}(\lambda-\lambda')\kappa_{j'}\log(1+e^{-\varepsilon_{j'}(\lambda')})\, .
\ee
Here the `effective energy' $\varepsilon_j$ was introduced to parametrize the filling fractions as
$\vartheta_j(\lambda)=(e^{\varepsilon_j(\lambda)}+1)^{-1}$.
This set of equations, together with Eq. \eqref{eq_STBA}, determines thermal states.
The spin Drude weight can be exactly computed within the theory of Generalized Hydrodynamics \cite{Doyon2018cor} and reads
\be
\mathcal{D}=\sum_{j=1}^N\int_{-\Lambda}^{\Lambda}\dd\lambda\, \rho_j(\lambda)(1-\vartheta_j(\lambda))(m^\dr_j v^\text{eff}_j(\lambda))^2\, ,
\ee
where $v^\text{eff}_j(\lambda)$ is the effective velocity of the quasiparticles renormalized by the background excitations, defined as  $v^\text{eff}_j(\lambda)=(\partial_\lambda e_j(\lambda))^\dr/(\partial_\lambda p_j(\lambda))^\dr$ where for each test function $g_j(\lambda)$ the dressing operation $g_j(\lambda)\to g_j^\dr(\lambda)$ is defined as
\be
g^\dr_j(\lambda)=g_j(\lambda)-\sum_{j'=1}^N\int_{-\Lambda}^{\Lambda}\int \dd\lambda\,  T_{j,j'}(\lambda-\lambda')\kappa_{j'}\vartheta_{j'}(\lambda')g_{j'}(\lambda')\, .
\ee

These general formulas need to be specified for the quantum magnet. In this case, one needs to distinguish the easy-axis regime $\varrho=\varrho_\text{ea}\in [0,+\infty)$ and the easy-plane regime $\varrho=i \varrho_\text{ep}$ with $\varrho_\text{ep}\in[0,\pi/(2S)]$. 
The isotropic point can be obtained as a limiting case of the easy-axis phase.
\bigskip

\textbf{The easy-axis regime.---} 
In this phase, the rapidity lives on a compact interval $\lambda\in[-\Lambda,\Lambda]=[-\pi/2,\pi/2]$, while the string's index is unbounded $N=\infty$.
One conveniently introduces the function
\be
K_j(\lambda)=\frac{1}{2\pi i}\partial_\lambda\log\left(\frac{\sin(\lambda-i j \varrho_\text{ea}/2)}{\sin(\lambda+i j \varrho_\text{ea}/2)}\right)\, .
\ee
Then the scattering kernel is computed as
\be
T_{j,j'}(\lambda)=\sum_{a=1}^{\min(j-1,j')}K_{|j-j'|-1+2a}(\lambda)+\sum_{\ell=1}^{\min(j+1,j')}K_{|j-j'|-1+2a}(\lambda)
\ee
The bare momentum, energy and magnetization are
\be
\frac{1}{2\pi}\partial_\lambda p_j(\lambda)=\sum_{a=1}^{\min(j,S)}K_{|j-j'|-1+2a}(\lambda)\, , \hspace{1pc}e_j(\lambda)=2\sinh\varrho_\text{ea} \partial_\lambda p_j(\lambda)\hspace{2pc} m_j=2j\, .
\ee
The parity is positive for all the strings $\kappa_j=1$.
In the easy-axis regime, there is an extra caveat concerning the magnetization. We are using the convention where excitations are placed on top of the all-spins-up reference state. By increasing the chemical potential, the magnetization covers the sector $[1,0]$, but it cannot change sign. The other sector is described by using the $\mathbb{Z}_2$ symmetry of the model under a global flip in the magnetization sign: in this way, excitations are placed on top of the all-spins-up reference state, and cover the other magnetization sector. The same feature is well-known in the canonical spin$-1/2$ XXZ chain \cite{Piroli2017}.
\bigskip

\textbf{The easy-plane regime.---} In the easy-plane, we parametrize the interaction as $\varrho=i\varrho_\text{ep}$. To guarantee the hermicity of the spin chain, interactions must be restricted to the region $\pi/\varrho_\text{ep}>2S$. Similarly to the spin$-1/2$ XXZ chain, the strings' content greatly depends on the continued-fraction representation of $\pi/\varrho_\text{ep}$. For the sake of simplicity, and since this is sufficient for taking the semiclassical limit, we focus on the so called root of units $\rhop=\pi/\ell$ with $\ell\in \mathbb{N}$.
In the planar regime, the rapidity lives on the whole real axis $\Lambda=\infty$, while the number of strings is finite $j\in\{1,...,\ell\}$. The magnetization and parity are respectively
\be
\begin{cases} m_j=2j \hspace{2pc} j<\ell\\
m_\ell=2\end{cases}\hspace{2pc}\begin{cases} \kappa_j=1 \hspace{2pc} j<\ell\\
\kappa_\ell=-1\end{cases}
\ee
We define the auxiliary function
\be
a_x^y=\frac{y}{\pi}\frac{\sin(\rhop x)}{\cosh(2\lambda)-y\cos(\rhop x)}\, .
\ee
Then, the scattering kernels can be written as
\be
T_{j,k}=(1-\delta_{m_j,m_k})a^{\sigma_j\sigma_k}_{|m_j-m_k|/2}(\lambda)+a^{\sigma_j\sigma_k}_{(m_j+m_k)/2}(\lambda)+2\sum_{\ell=1}^{\min(m_j,m_k)/2-2}a^{\sigma_j\sigma_k}_{|m_j-m_k|/2+2\ell}\, .
\ee
The bare energy is $e_j(\lambda)=2\sin\rhop\partial_\lambda p_j(\lambda)$, while the $\lambda-$derivative of the bare momentum is
\be
p_j(\lambda)=2\pi\sum_{k=1}^{\min(m_j,2S)} a^{\sigma_j}_{|m_j/2-2S|+2k-1}(\lambda)\, .
\ee

In the high-temperature limit, $\beta \to 0$, the TBA equations are rendered algebraic, depending only on the chemical potential $\mu$. The occupation function, valid for any spin $S$, take the following explicit form \cite{takahashi2005thermodynamics}:
\be\label{eq_S_infQEP}
\vartheta_{j\le\ell-2}=\frac{\sinh^2(\mu)}{\sinh^2(\mu(j+1))}\hspace{1pc}\vartheta_{\ell-1}=\frac{1}{1+e^{\ell \mu}\frac{\sinh((\ell-1)\mu)}{\sinh(\mu)}}\hspace{1pc}\vartheta_{\ell}=\frac{1}{1+e^{\ell \mu}\frac{\sinh(\mu)}{\sinh((\ell-1)\mu)}}
\ee

\section{The spin-wave limit of the classical spin chain}
\label{sec_SW}

In the limit of strong polarization $S_j^3\sim 1$, the equation of motion can be linearized or, equivalently, the Hamiltonian can be expanded up to quadratic order in the fluctuations around the polarized state. In this limit, the excitations are spin waves, i.e. non-interacting plane waves with Rayleigh-Jeans statistics whose thermodynamics can be easily computed. Since this limiting case is an important benchmark for our exact thermodynamic description, it is worth discussing the relevant details.
We consider the lattice Hamiltonian in the easy-axis regime (2) and leave arbitrary the normalization of the Lie-Poisson brackets $\Sp$, i.e. $\{S^a,S^b\}=\Sp\epsilon_{abc}S^c$. By looking at the equation of motion $\frac{\dd S_j^a}{\dd t}=\Sp \epsilon_{abc}\frac{\partial H_{LLL}}{\partial S^b_j}S^c_j$, and expanding up to linear order in $S_j^1$ and $S_j^2$, we reach the approximated equation of motion
\be
\frac{\dd S^+(k)}{\dd t}\simeq-i \omega_\text{SW}(k)S^+(k)\hspace{2pc}\omega_\text{SW}(k)=\Sp\frac{2\rhoa}{\sinh\rhoa}\big(\cosh\rhoa-\cos k\big)
\ee
where we define $S^\pm(k)=\sum_j e^{-ik j}\frac{1}{\sqrt{2}}(S^1_j+iS^2_j)$. Importantly, the group velocity of the spin-wave is given by $v_\text{SW}(k)=\partial_k \omega_\text{SW}(k)$.
One could be tempted to identify the frequency $\omega_\text{SW}(k)$ with the energy of the spin-wave $e_\text{SW}(k)$, but this is not the case: thermodynamic quantities, such as the expectation value of the Hamiltonian for example, are not $\Sp-$dependent, while $\omega_\text{SW}(k)$ is.

It turns out that the energy and the frequency differ for a simple $\Sp-$dependent proportionality factor $\omega_\text{SW}(k)=C(\Sp)e_\text{SW}(k)$: the quickest way to fix $C(\Sp)$ is looking at the Hamiltonian expectation value in the spin-wave limit. On the one hand, we use the thermodynamics of spin-wave: since they are non-interacting plane waves, their mode-density $n(k)$ follows Rayleigh-Jeans distribution $n(k)=1/(\beta e_\text{SW}(k)+2\mu)$, thus
\be
\langle H_\text{LLL}\rangle \simeq \langle H_\text{LLL}\rangle_\text{vac}+L\int \frac{\dd k}{2\pi}\frac{e_\text{SW}(k)}{\beta e_\text{SW}(k)+2\mu}\, 
\ee
where $\langle H_\text{LLL}\rangle_\text{vac}$ is the expectation value of the Hamiltonian on the fully polarized state.
In particular, on infinite temperature states $\beta=0$ we find $\langle H_\text{LLL}\rangle \simeq \langle H_\text{LLL}\rangle_\text{vac}+L\frac{\Sp}{C(\Sp)}\frac{\rhoa \coth\rhoa}{\mu}$. On the other hand, we can expand the Hamiltonian for small fluctuations around the polarized state
\be
H_\text{LLL}\simeq \langle H_\text{LLL}\rangle_\text{vac}+\sum_j\frac{2\rhoa}{\sinh\rhoa}\left(\frac{\cos\rhoa}{4}\left(S^1_{j}S^1_{j}+S^2_{j}S^2_{j}+S^1_{j+1}S^1_{j+1}+S^2_{j+1}S^1_{j+1}\right)-\frac{1}{2}\left(S^1_jS^1_{j+1}+S^2_jS^2_{j+1}\right)\right)\, .
\ee
On infinite temperature states, spins at different positions are uncorrelated, thus $\langle S^a_j S^b_{j+1}\rangle=\langle S^a_j \rangle\langle S^b_{j+1}\rangle$, while for large $\mu$ one readily computes $\langle S^1_{j}S^1_{j}+S^2_{j}S^2_{j}\rangle\simeq \frac{2}{\mu}$, thus finding $H_\text{LLL}\simeq \langle H_\text{LLL}\rangle_\text{vac}+L\frac{2\rhoa\coth\rhoa}{ \mu}$: imposing the consistency with the spin-wave thermodynamics, we therefore fix $C(R)=\frac{\Sp}{2}$ and thus
\be
e_\text{SW}(k)=\frac{4\rhoa}{\sinh\rhoa}\big(\cosh\rhoa-\cos k\big)\, .
\ee
Notice that only in the case where $\Sp=2$, the spin-wave group velocity is the momentum derivative of the energy $v_\text{SW}(k)=\partial_k e_\text{SW}(k)$, which is the canonical convention in GHD.
We notice the spin-wave analysis of the easy-plane is readily obtained by analytical continuation.

\section{The semiclassical limit of the quantum TBA}
\label{sec_semiclassical}

In this section, we provide the details of the semiclassical limit of the TBA of the integrable quantum magnet, summarized in Section \eqref{sec_quantum}. 
We performed the semiclassical limit at the level of TBA. To make sure the limit yields the lattice Hamiltonian given by Eq. (2) (with the Lie-Poisson bracket normalization $\Sp=2$), we have performed several consistency checks. In particular, (i) the spin wave limit is analytically retrieved, (ii) at the isotropic point we agree with Ref. \cite{Haldane1982}, iii) in the continuum limit, the dispersion relations and scattering kernels match the well-known expression for the Landau-Lifschitz field theory, (iv) the large-spin limit of the fundamental commutation (RLL) relation yields the (quadratic) Sklyanin algebra associated with the axially anisotropic lattice Landau--Lifshitz model (not reported), and finally (v) our analytics complies with great accuracy with ab initio numerical simulations.

As we discussed in the main text, the semiclassical limit is achieved by combining a large spin limit $S\to \infty$ with a small interaction limit. For the sake of clarity, we analyze separately the easy-axis and easy-plane regimes. We report the equations for the isotropic point and finally discuss the continuum limit.
Within this section, we add a label ``q" to quantities belonging to the quantum model.

\subsection{The easy-axis regime}
\label{subsec_easy_axis}

We use the quantum spin $S$ as a control parameter and rescale the quantum interaction as $\rhoa=(2S)\varrho_{\text{ea};q}$: we will check that this rescaling gives the correct interaction in the classical model by looking at the spin-wave limit.
The semiclassical limit of the easy-axis regime closely follows the analysis of the attractive Non-Linear Schr\"{o}dinger \cite{Koch2022} and sine-Gordon model \cite{Koch2023}. Further details can be found in these references, here we focus on the main points.
Finite TBA equations are obtained by assuming the strings can be replaced by a continuous spectral parameter $\sigma$ with the correspondence $2S\sigma\leftrightarrow j$.
Using this definition, in the large $S$ limit one obtains
\be\label{eq_STaxis}
T_{q;j,j'}(\lambda)\to 2 S T_{\sigma,\sigma'}(\lambda)\, ,\hspace{2pc}T_{\sigma,\sigma'}(\lambda)=\frac{1}{\pi\rhoa}\log\left(\frac{\cosh((\sigma+\sigma')\rhoa)-\cos(2\lambda)}{\cosh((\sigma-\sigma')\rhoa)-\cos(2\lambda)}\right)\, .
\ee
Notice the useful identity $\int_{-\pi/2}^{\pi/2}\dd\lambda\, T_{\sigma,\sigma'}(\lambda)=2\min(\sigma,\sigma')$.
The root density of each quantum string vanishes as $\rho_{q;j}\to \frac{1}{2S}\rho_\sigma(\lambda)$: the system accommodates for the growing magnetization by populating more strings. In contrast, the quantum total root density diverges $\rho^t_{q;j}(\lambda)\to 2S \rho^t_\sigma(\lambda)$. Indeed, the bare momentum and magnetization diverge as $\partial_\lambda p_{q;j}\to 2S \partial_\lambda p_{\sigma}(\lambda)$ and $m_{q;j}\to 2S m_\sigma$, while the energy has a finite limit $e_{q;j}\to e_{\sigma}(\lambda)$ and the parity is positive for all modes $\kappa_\sigma=1$.
We define $\partial_\lambda p_\sigma(\lambda)=\pi T_{\sigma,1}(\lambda)$, $e_\sigma(\lambda)=2\rhoa\partial_\lambda p_\sigma(\lambda)$ and $m_\sigma=2\sigma$.
We finally look at the limit of the quantum Yang-Yang entropy: here, extra caveat should be taken due to apparently divergent terms
\be\label{eq_Slimit}
\mathcal{S}_q\to L\int_{\delta_S}^\infty\dd\sigma \int_{-\pi/2}^{\pi/2}\dd\lambda \left[\rho_\sigma (\lambda)-\rho_\sigma (\lambda)\log(\vartheta_\sigma(\lambda))+2\log(2S)\rho_\sigma(\lambda)\right]\, .
\ee
Above, we introduced the classical filling $\vartheta_\sigma(\lambda)=\rho_\sigma(\lambda)/\rho_\sigma^t(\lambda)$. In the integrand, a logarithmic divergent term appears: this must be compensated by another divergent counterterm that arises from small strings. For this reason, we introduce a cutoff $\delta_S\to 0$: its value must be fixed in a self-consistent manner, by requiring the final TBA equations obtained by minimizing the free energy are finite. Only after this, the limit $S\to \infty$ is safely taken. This procedure has been introduced in Refs. \cite{Koch2022,Koch2023}: following the same steps, one needs to impose $\log(2S \delta_S)=1$ and finally taking the $S\to\infty$ limit one reaches the set of TBA equations
\be\label{eq_STBA}
\sigma^2\varepsilon_\sigma(\lambda)=\beta e_\sigma(\lambda)+\mu m_\sigma +\int_0^{\infty} \dd\sigma'\dd\lambda' T_{\sigma,\sigma'}(\lambda-\lambda')\frac{e^{-(\sigma')^2\varepsilon_{\sigma'}(\lambda')}-1}{(\sigma')^2}\, .
\ee
where the effective energy $\varepsilon_\sigma$ parametrizes the filling function as $\vartheta_\sigma(\lambda)=\sigma^{-2}e^{-\sigma^2\varepsilon_\sigma(\lambda)}$. At small values of $\sigma$, $\varepsilon_\sigma$ reaches a finite value, hence the filling function develops a $\sim 1/\sigma^2$ divergence.
As in the quantum XXZ chain \cite{Piroli2017}, there are two disconnected sets of TBA equations corresponding to a two-fold degeneracy of the vacuum state, requiring and extra $\mathbb{Z}_{2}$ label for specifying the corresponding sector of positive or negative density of magnetization.

\begin{figure}[t!]
\centering
\includegraphics[width=0.95\textwidth]{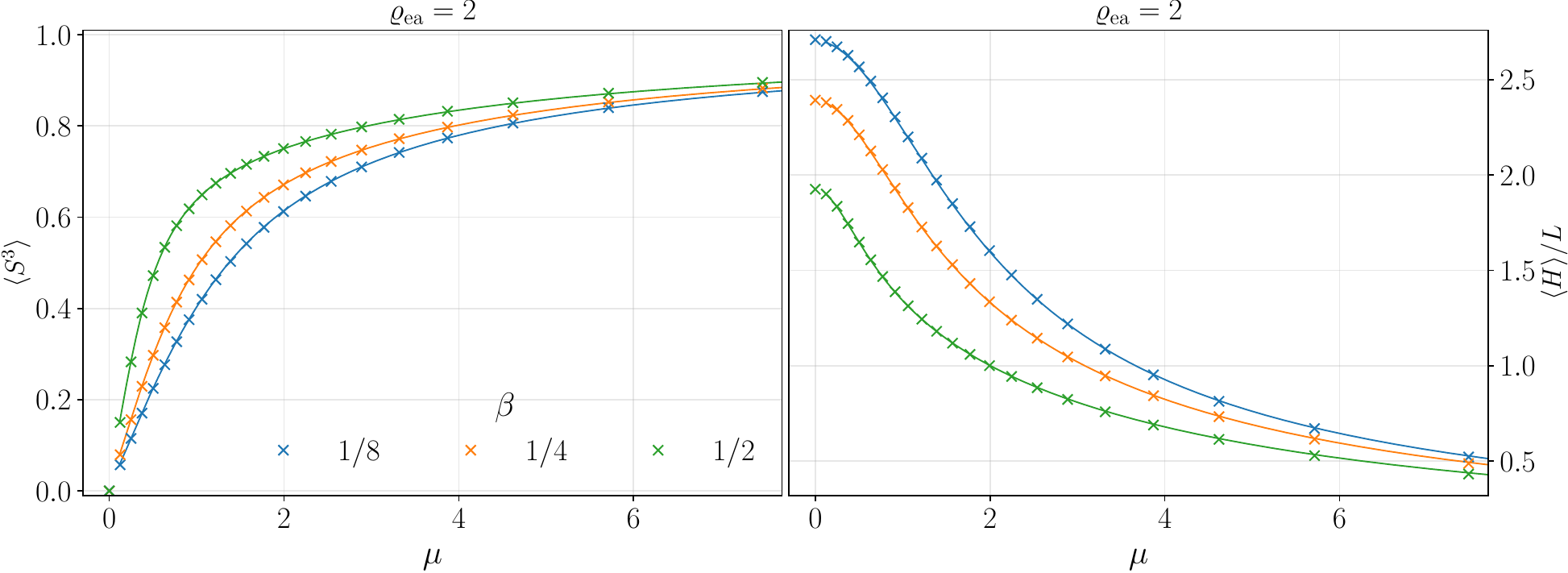}
 \caption{Average magnetization (left) and energy (right) obtained by solving classical TBA equations (solid lines) compared with Monte-Carlo simulation (crosses) in the easy-axis regime with $\rho_{\rm ea}=2$ for $\beta \in \{1/8, 1/4, 1/2\}$. Monte Carlo error bars are negligible on the plot's scale and thus omitted.}   
	\label{fig_S1}
\end{figure}

The semiclassical limit straightforwardly extends to the definition of the dressing operation $g_\sigma(\lambda)\to g^\dr_\sigma(\lambda)$ as
\be\label{eq_SA_dr}
g_\sigma^\dr(\lambda)=g_\sigma(\lambda)-\int_{0}^\infty\dd\lambda'\, T_{\sigma,\sigma'}(\lambda-\lambda')\vartheta_{\sigma'}(\lambda')g_{\sigma'}^\dr(\lambda')\, ,
\ee
and to the spin Drude weight
\be\label{eq_Drude}
\mathcal{D}=\int_0^\infty\dd\sigma \int_{-\pi/2}^{\pi/2}\dd\lambda\rho_\sigma(\lambda)[m^\dr_\sigma(\lambda) v^\text{eff}_\sigma(\lambda)]^2\, ,
\ee
with $v^\text{eff}_\sigma(\lambda)=(\partial_\lambda e_\sigma)^\dr/(\partial_\lambda p_\sigma)^\dr$.
At large temperatures, the Drude weight diverges due to the singular dressed velocity $\lim_{(\sigma,\lambda) \to (1,0)}|v^\eff_\sigma(\lambda)|=+\infty$ inherited by the singularities of the bare energy and momentum. To extract the singularity, we proceed as follows.
We first change variable from the rapidity to energy $E$ by imposing $E=e_\sigma(\lambda)$: integrals will be carried out in the space $(\sigma,E)$. Being mostly interested in thermal states, we use the fact that the integrand in the Drude weight is symmetryc $\lambda\to -\lambda$, hence focus on positive $\lambda$. The singularity at $(\sigma,\lambda)=(1,0)$ is mapped into $E\to \infty$, hence we split the Drude weight in two contributions
\be\label{eq_S19}
\mathcal{D}=\text{finite part}+2\int_{0}^\infty\dd\sigma \int_{E_c}^{\infty}\frac{\dd E}{2\pi} \left[\vartheta_\sigma(\lambda) \frac{(\partial_\lambda p_\sigma)^\dr}{\partial_\lambda e_\sigma}[m^\dr_\sigma(\lambda) v^\text{eff}_\sigma(\lambda)]^2\right]_{E=e_\sigma(\lambda)}
\ee
Above, $E_c$ is some large, but otherwise arbitrary, cutoff.
At large $E$, we can extract from the effective energy $\varepsilon_\sigma(\lambda)$ its singular part, by defining $\varepsilon_{\text{NS}}=\lim_{(\sigma,\lambda) \to (1,0)}(\varepsilon_\sigma(\lambda)-\beta \sigma^{-2}e_\sigma)$, which is finite. Furthermore, the divergence in the effective velocity is due to the bare energy and momentum, hence we can approximate $v_\sigma^\eff(\lambda)\to (\partial_\lambda e_\sigma)/(\partial_\lambda p_\sigma)$. In the same spirit, we expand the energy $e_\sigma(\lambda)$ around the singularity and obtain the approximate change of variables $\frac{1}{2} (2\lambda)^2+\frac{1}{2}\rhoa^2(1-\sigma)^2\simeq e^{-E/2}(\cosh(2\rhoa)-1)$. Plugging these approximations in Eq. \eqref{eq_S19} and taking the integral over $\sigma$, at the leading order we obtain
\be\label{eq_asympD}
\mathcal{D}=\text{finite part}+\frac{1}{2}(m^{\text{dr}}_1(0))^2 e^{-\varepsilon_{NS}}\int_{E_c}^{\infty}\dd E\, \frac{e^{-\beta E}}{E} \, .
\ee
The last integral has a logarithmic singularity $\sim -\log\beta$, showing the divergence of the Drude weight in the infinite temperature limit.
In Fig. \ref{fig_S1}, we benchmark the TBA equations by computing the expectation value of the magnetization and energy on thermal states and comparing with ab-initio Monte Carlo simulations, finding excellent agreement.

\bigskip
\textbf{The infinite-temperature limit.---} It is worth focusing on the infinite temperature limit by sending $\beta\to 0$ in Eq. \eqref{eq_STBA}. The integral equation now becomes translational invariant in the rapidity space, therefore the effective energy becomes constant in the rapidity space $\varepsilon_\sigma(\lambda)\to \varepsilon_\sigma$. By averaging over $\lambda$, we can integrate the scattering kernel using its normalization, and eventually reaching the simplified equations
\be\label{eq_easyP_TBA_infiniteT}
\sigma^2\varepsilon_\sigma=2\mu\sigma+\int_0^\infty \dd\sigma' \, 2\min(\sigma,\sigma')\frac{e^{-(\sigma')^2\varepsilon_{\sigma'}}-1}{(\sigma')^2}\, .
\ee
This integral equation is formally identical to that faced in extracting the low-temperature behavior of the sine-Gordon field theory \cite{Koch2023} and can be analytically solved: we leave to the quoted reference a more detailed discussion, here we report the final result, yielding $e^{-\sigma^2\varepsilon_\sigma}=\frac{(\mu\sigma)^2}{\sinh^2(\mu\sigma)}$. The exact filling fraction can be used in the dressing equations \eqref{eq_SA_dr}. Let us assume that the test function $g_\sigma(\lambda)$ has the the following properties: \emph{i)} it is constant over the rapidities and \emph{ii)} it is linear in $\sigma$, i.e. $g_\sigma(\lambda)=\bar{g}\sigma$. Then, the dressing equation can be analytically solved $g_\sigma^\dr=\bar{g} u(2\mu\sigma)/(2\mu)$ with $u(x)=x\coth(x/2)-2$ \cite{Koch2023}.

These assumptions are verified by the bare magnetization, hence the dressed magnetization $m_\sigma^\dr$ can be computed exactly on infinite temperature thermal states. One can finally compute the total magnetization as
\be\label{eq_S_infT}
\langle S^z\rangle=1-\int_{-\pi/2}^{\pi/2} \dd\lambda \int_0^\infty \dd\sigma\, m_\sigma \rho_\sigma(\lambda)=1-\int_{-\pi/2}^{\pi/2}  \frac{\dd\lambda}{2\pi} \int_0^\infty \dd\sigma\, m_\sigma \vartheta_\sigma(\lambda)(\partial_\lambda p_\sigma)^\dr=1-\int_{-\pi/2}^{\pi/2} \frac{\dd\lambda}{2\pi} \int_0^\infty \dd\sigma \,m^\dr_\sigma \vartheta_\sigma(\lambda)(\partial_\lambda p_\sigma)\, .
\ee
Since the dressed magnetization and the filling fraction on infinite-temperature states are rapidity-independent,  the integral over $\partial_\lambda p_\sigma$ can be factorized $\int \dd\lambda \frac{\partial_\lambda p_\sigma}{2\pi}=\min(\sigma,1)$.
The last integral over $\sigma$ can be easily computed, giving $\langle S^z\rangle=-\frac{1}{\mu}+\coth\mu$ which matches with the ab initio computation from the microscopic model.
In the last passage Eq. \eqref{eq_S_infT}, we use the symmetry $\int_{-\pi/2}^{\pi/2} \dd\lambda\int_0^\infty\dd\sigma\, \vartheta_\sigma(\lambda) g^\dr_\sigma(\lambda) h_\sigma(\lambda)=\int_{-\pi/2}^{\pi/2}  \dd\lambda\int_0^\infty\dd\sigma\, \vartheta_\sigma(\lambda) g_\sigma(\lambda) h^\dr_\sigma(\lambda)$ that holds for any test function $g_\sigma(\lambda)$ and $h_\sigma(\lambda)$.

\bigskip
\textbf{The classical entropy.---} It is expected that the limit of the quantum entropy \eqref{eq_Slimit} should describe the classical entropy, but dealing with the apparent $\log S$ divergences seems challenging. Therefore, we use another route: we postulate the entropy can be written as $\mathcal{S}=L\int \dd\sigma \int \dd\lambda \rho_\sigma^t s_\sigma(\vartheta_\sigma)+L C$ with a proper entropy weight $s_\sigma$ and possibly a constant $C$. Then, we consistently require that the minimization of the free energy $F=\beta \langle H\rangle+\mu \langle M^3-M^3_\text{vac}\rangle-\mathcal{S} $ gives Eq. \eqref{eq_STBA}. After straightforward manipulations, we obtain $s_\sigma(\vartheta_\sigma)=\vartheta_\sigma-1/\sigma^2-\log(\sigma^2\vartheta_\sigma)\vartheta_\sigma$.
To fix the constant $C$, we compute the free energy $F=L\int_0^\infty\dd\sigma\int \dd\lambda \frac{\partial_\lambda p_\sigma}{2\pi}\mathcal{F}_\sigma(\vartheta_\sigma)-L C$ with $\mathcal{F}_\sigma(\vartheta_\sigma)=\vartheta_\sigma s'_\sigma(\vartheta_\sigma)-s_\sigma(\vartheta_\sigma)=\frac{1}{\sigma^2}-\vartheta_\sigma(\lambda)$: using that for $\beta=0$ the filling is independent over $\lambda$, we carry out the rapidity integral over $\partial_\lambda p_\sigma$ hence we obtain
\be
F=L\int_{0}^\infty\dd\sigma \frac{\min(\sigma,1)}{\sigma^2}\left(1-\frac{(\mu\sigma)^2}{\sinh^2(\mu\sigma)}\right)-LC=L(\mu+\log(\mu/\sinh\mu)-C)\, .
\ee
On the other hand, the partition function of a classical spin in a magnetic field and with flat measure on the unit sphere is $Z=\frac{1}{4\pi}\int_{0}^\pi \dd\phi\int_{0}^{2\pi}\dd\theta\sin\phi  \, e^{\mu\cos\phi-\mu}=e^{-\mu}\frac{\sinh\mu}{\mu}$, hence $F=- L \log Z$ and finally, by comparison, $C= 0$.

\bigskip
\textbf{The spin-wave limit.---} We now focus on the strongly polarized limit $\mu\gg 1$, and retrieve from the exact TBA the spin-wave thermodynamics outlined in Section \ref{sec_SW}.
spin-wave are radiative in nature and should be recovered by the solitonic TBA: we closely follow the same procedure used in studying the low-temperature limit of the sine-Gordon field theory \cite{Koch2023}.

In the limit $\mu\gg 1$, only solitons with small $\sigma$ are populated. In this limit, the kernel $T_{\sigma,\sigma'}(\lambda)$ is increasingly peaked at small rapidities and can be eventually approximated with a delta function $T_{\sigma,\sigma'}(\lambda)\simeq2\min(\sigma,\sigma')\delta(\lambda)$ for $\sigma\ll 1$ and $\sigma'\ll 1$. By plugging this approximation in Eq. \eqref{eq_STBA} and furthermore approximating $e_\sigma(\lambda)\simeq \sigma e^{(1)}(\lambda)$, with $e^{(1)}(\lambda)=\lim_{\sigma\to 0}\sigma^{-1}e_\sigma(\lambda)$, one obtains
\be\label{eq_A_lowT}
\sigma^2\varepsilon_\sigma(\lambda)=\beta e^{(1)}(\lambda)+2\mu\sigma+\int_0^\infty \dd\sigma' 2\min(\sigma,\sigma')\frac{e^{-(\sigma')^2\varepsilon_{\sigma'}(\lambda)}-1}{(\sigma')^2}\, .
\ee
This equation is formally identical to Eq.\eqref{eq_easyP_TBA_infiniteT}, with a parametric dependence on the rapidity, hence it can be solved as before. The same strategy can be used for the dressing equations.
To recover radiation, it is instructive looking at the expectation value of a charge $q_\sigma(\lambda)$ on a low temperature state. In particular, we fix the rapidity and look at the cumulative effects of solitons (see Ref. \cite{Koch2023} for details)
\be\label{eq_soliton_resumming}
\int_0^\infty \dd\sigma \rho_\sigma q_\sigma(\lambda)\simeq \frac{1}{2\pi}\frac{\partial_\lambda p^{(1)}(\lambda)}{\beta e^{(1)}(\lambda)+\mu}q^{(1)}(\lambda)\, .
\ee
The r.h.s. contribution is formally identical to the thermal contribution of a radiative mode, hence obeying the Rayleigh-Jeans distribution, with energy $e^{(1)}(\lambda)$, momentum derivative $\partial_\lambda p^{(1)}(\lambda)=\lim_{\sigma\to 0}\sigma^{-1}\partial_\lambda p_{\sigma}(\lambda)$ and carrying a charge $q^{(1)}(\lambda)=\lim_{\sigma\to 0}\sigma^{-1}q_{\sigma}(\lambda)$. 
To push further this identification, we define the momentum $k\to k(\lambda)=\int^\lambda \dd\lambda'\partial_{\lambda'}p^{(1)}(\lambda')=2\arctan(\tan\lambda\coth(\rhoa/2))+\pi$. By inverting this relation $\lambda(k)$ and plugging in the energy $e^{(1)}(\lambda)$, we obtain the energy of spin waves $e_\text{SW}(k)=e^{(1)}(\lambda(k))=\frac{4\rhoa}{\sinh\rhoa}(\cosh\rhoa-\cos k)$ and thus we match the approximate thermodynamics of small fluctuations nearby the ferromagnetic vacuum discussed in Section \ref{sec_SW}. Notice that, furthermore, in the spin wave limit we readly find that $v^\text{eff}_\sigma(\lambda)\to \partial_k e_\text{SW}(k)$, thus fixing to $R=2$ the normalization in the Lie-Poisson brackets, according to the argument presented in Section \ref{sec_SW}.

\subsection{The easy-plane regime}
\label{subsec_easy_plane}
We now discuss in detail the semiclassical limit of the easy-plane regime. As in the previous case, we rescale the interaction as $\rhop=(2S)\varrho_{\text{ep};q}$ and take the limit $S\to \infty$. As discussed in Section \ref{sec_quantum}, the TBA of the quantum chain depends on the interactions in a seemingly non-continuous way. However, this structure is washed away in the semiclassical limit, therefore for the sake of simplicity we can consider the so called roots of unity $\varrho_{\text{ep};q}=\pi/\ell_q$ with $\ell_q$ a large integer. For convenience, we define $\ell=\ell_q/(2S)$: notice $\ell$ remains constant in the semiclassical limit.
In the quantum chain there are $\ell_q$ strings which behave differently in order to achieve a finite semiclassical limit: the strings $j\in\{1,...,\ell_q-2\}$ behave alike the easy-axis case and can be well-described by solitons with a continuum spectral parameter $\sigma\in[0,\ell]$, with the correspondence $\sigma= j/(2S)$. For these modes, we borrow the same notation as the easy-axis regime. The two last strings behave differently: the last string contributes as a radiative mode denoted with ``$\R$", and its scaling is similar to semiclassical limits of field theories with only radiation \cite{Bastianello2018}. In contrast, the second last describes solitons with no bare energy or momentum, but maximum magnetization, and we conventionally denote them with ``$\M$".
In short, the filling functions and root densities scale as
\be
\vartheta_{q;j<\ell_q-1}(\lambda)\to \frac{1}{(2S)^2}\vartheta_{\sigma}(\lambda)\, ,\hspace{1pc}\vartheta_{q;\ell_q-1}(\lambda)\to \frac{1}{2S}\vartheta_{\M}(\lambda)\, ,\hspace{1pc}\vartheta_{q;\ell_q}(\lambda)=1-\frac{1}{2S}\vartheta^{-1}_\R(\lambda)\, .
\ee
\be
\rho_{q;j<\ell_q-1}(\lambda)\to \frac{1}{2S}\rho_{\sigma}(\lambda)\, ,\hspace{1pc}\rho_{q;\ell_q-1}(\lambda)\to \rho_{\M}(\lambda)\, ,\hspace{1pc}\rho_{q;\ell_q}(\lambda)=2S\rho_\R(\lambda)\, .
\ee
while the quantum kernels behave as
\begin{eqnarray}
T_{q;j<\ell_q-1,j'<\ell_q-1}(\lambda)\to 2S T_{\sigma,\sigma'}(\lambda)\, ,&&\hspace{1pc} T_{q;j<\ell_q-1,\ell_q}(\lambda)\to T_{\sigma,\R}(\lambda)\, ,\\T_{q;\ell_q-1,\ell_q}(\lambda)\to T_{\M,\R}\,,&&\hspace{1pc} T_{q;\ell_q,\ell_q}(\lambda)\to\delta(\lambda)+\frac{1}{2S}T_{\R,\R}(\lambda)\, ,
\end{eqnarray}
with
\begin{eqnarray}
T_{\sigma,\sigma'}(\lambda)=\frac{1}{\rhop\pi}\log\left[\frac{\cosh(2\lambda)-\cos(\rhop(\sigma+\sigma'))}{\cosh(2\lambda)-\cos(\rhop(\sigma-\sigma'))}\right]\, ,&&\hspace{1pc}T_{\sigma,\R}(\lambda)=T_{\R,\sigma}(\lambda)=-\frac{2}{\pi}\frac{\sin(\rhop\sigma)}{\cosh(2\lambda)+\cos(\rhop\sigma)}\, ,\\
T_{\R,\R}(\lambda)=-\frac{\rhop}{2\pi}\partial_\lambda \Pv\coth(\lambda)\, ,&&\hspace{1pc}T_{\M,\R}(\lambda)=T_{\R,\M}(\lambda)=-\delta(\lambda)\, .
\end{eqnarray}
Omitted classical kernels vanish in the limit. The self-interaction of radiation $T_{\R,\R}$ needs further explanation: the double pole must be regularized within the principal value regularization, denoted with $\Pv$ such as 
\be
\int_{-\infty}^{\infty} \dd\lambda'\, T_{\R,\R}(\lambda-\lambda')g_{\R}(\lambda')=\left[\frac{\rhop}{2\pi}\coth[2(\lambda'-\lambda)]g_{\R}(\lambda')\right]_{\lambda'=-\infty}^{\lambda=\infty}+\lim_{\epsilon\to 0^+}\int_{\lambda'\in[\infty,-\epsilon]\cup[\epsilon,+\infty]}\dd\lambda'\,\frac{\rhop}{2\pi}\coth[2(\lambda-\lambda')]\partial_{\lambda'}g_{\R}(\lambda')\, ,
\ee
for any test function $g_{\R}(\lambda)$. It is worth noticing the normalizations 
\be
\int\dd\lambda T_{\sigma,\sigma}(\lambda)=2\min(\sigma,\sigma')-2\sigma \sigma'/\ell\, , \hspace{1pc}\int \dd\lambda T_{\sigma,\R}(\lambda)=-2\sigma/\ell\, , \hspace{1pc}\int \dd\lambda T_{\R,\R}(\lambda)=-2/\ell\, .
\ee

Verifying that this scaling gives finite dressing equations in the $S\to\infty$ limit is a lengthy, but straightforward, derivation.
When computing the semiclassical limit of the Yang-Yang entropy, divergent terms akin to the easy-axis regime are obtained
\be
\mathcal{S}_{q}\to L\int_{-\infty}^{\infty}\dd\lambda'\left\{\int_{\delta_S}^\ell \dd\sigma\, \rho_\sigma(1-\log\vartheta_\sigma)+\rho^t_\R(\log\vartheta_R+1)+\rho\M(1-\log\vartheta_\M)+\log(2S)\left[\rho_\M^t+\rho_\M+\int_{\delta_S}^{\ell}\dd\sigma \, 2\rho_\sigma\right]\right\}\, .
\ee
The explicit $\log(2S)$ divergence is counterbalanced by a divergence at small $\sigma$ regularized by $\delta_S$: finite TBA equations can be obtained if one imposes again $\log(2S \delta_S)=1$, leading to
\be\label{eq_TBA_P1}
\sigma^2\varepsilon_\sigma(\lambda)=\beta \sigma e_\sigma(\lambda)+\mu m_\sigma-\frac{2\sigma}{\ell}\log\ell+\int_0^\ell \dd\sigma'\int\dd\lambda\, T_{\sigma,\sigma'}(\lambda-\lambda') \frac{e^{-(\sigma')^2\varepsilon_{\sigma'}(\lambda')}-1}{(\sigma')^2}+\int\dd\lambda\, T_{\sigma,\R}(\lambda-\lambda') \log\varepsilon_\R(\lambda')\, ,
\ee
\begin{multline}\label{eq_TBA_P2}
\varepsilon_\R(\lambda)=\beta e_\R(\lambda)+\mu m_\R+\frac{2}{\ell}-\frac{2}{\ell}\log\ell+\\\int_0^\ell \dd\sigma'\int\dd\lambda\, T_{\R,\sigma'}(\lambda-\lambda') \frac{e^{-(\sigma')^2\varepsilon_{\sigma'}(\lambda')}-1}{(\sigma')^2}+\int\dd\lambda\, T_{\R,\R}(\lambda-\lambda') \log\varepsilon_\R(\lambda')+\int\dd\lambda\, T_{\R,\M}(\lambda-\lambda') e^{-\varepsilon_\M(\lambda')}\, ,
\end{multline}
\be\label{eq_TBA_P3}
\varepsilon_{\M}(\lambda)=\mu m_\M+\int\dd\lambda\, T_{\R,\M}(\lambda-\lambda')\log\varepsilon_\R(\lambda')\, .
\ee
The bare magnetizations are respectively $m_\sigma=2\sigma$, $m_\M=2\ell$, $m_\R=2$, while the bare energies are $e_\sigma=2\rhop\partial_\lambda p_\sigma$, $e_\R=2\rhop\partial_\lambda p_\R$ and $e_\M=2\rhop\partial_\lambda p_\M$, where
\be
\partial_\lambda p_\sigma(\lambda)=\pi T_{\sigma,1}(\lambda)\, ,\hspace{1pc}\partial_\lambda p_\R=-\frac{2\sin\rhop}{\cosh(2\lambda)+\cos\rhop}\, ,\hspace{1pc}\partial_\lambda p_\M=0\, .
\ee
The classical modes inherit the parity of the quantum ones, so $\kappa_\sigma=1$, $\kappa_\M=1$ and $\kappa_\R=-1$.
Above, the effective energies parametrize the filling functions as
\be
\vartheta_\sigma(\lambda)=\frac{e^{-\sigma^2\varepsilon_\sigma(\lambda)}}{\sigma^2}\, ,\hspace{1pc}\vartheta_\R(\lambda)=\frac{1}{\varepsilon_\R(\lambda)}\, , \hspace{1pc}\vartheta_\M(\lambda)=e^{-\varepsilon_\M(\lambda)}\, .
\ee
Unlike in the easy-axis regime a single set of TBA equations suffices to capture states from both magnetization sectors by tuning the chemical $\mu \in \mathbb{R}$. 

\begin{figure}[t!]
\centering
\includegraphics[width=0.95\textwidth]{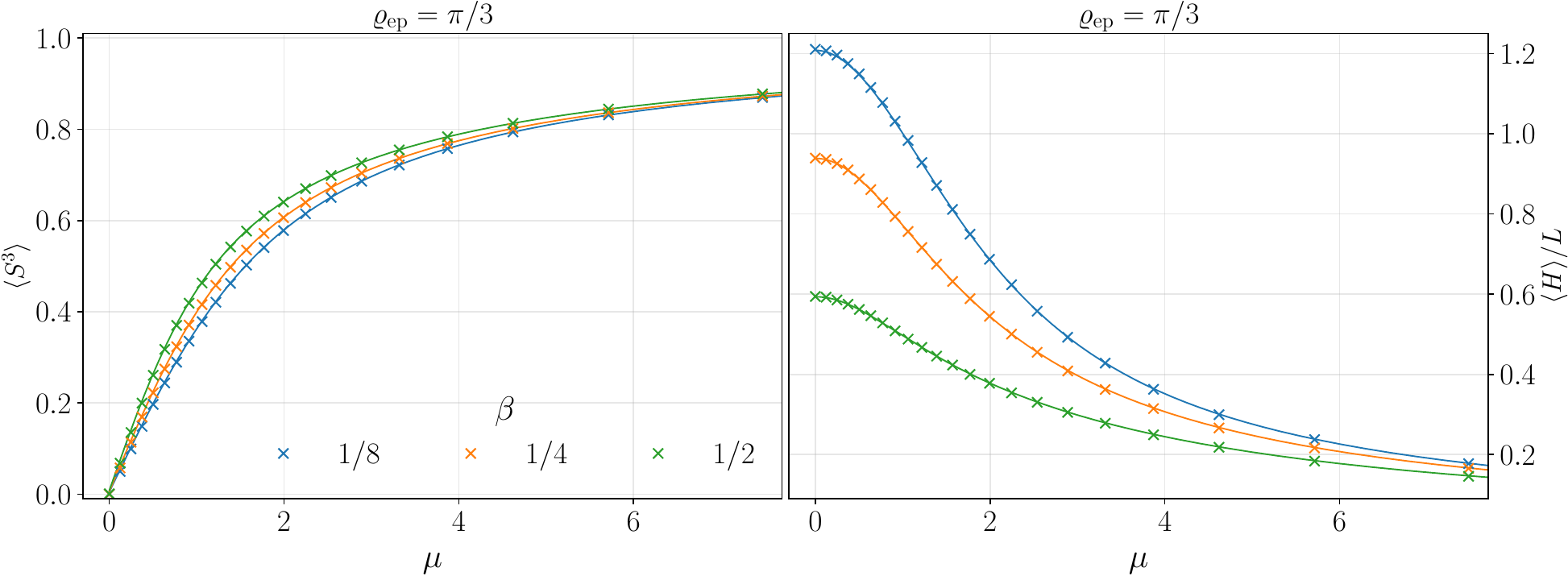}
 \caption{Average magnetization (left) and energy (right) obtained by solving classical TBA equations (solid lines) compared with Monte-Carlo simulation (crosses) in the easy-plane regime with $\rho_{\rm ea}=\pi/3$ for $\beta \in \{1/8, 1/4, 1/2\}$. Monte Carlo error bars are negligible on the plot's scale and thus omitted.}  
	\label{fig_S2}
\end{figure}

The dressing operation has the canonical form where the sum over all the particle species should be taken into account, weighted with the proper scattering kernels.
Finally, we focus on the Drude weight: by taking the semiclassical limit of the quantum expression, one readily reaches
\be
\mathcal{D}=\int\dd\lambda\int_0^\ell \dd\sigma\, \rho_\sigma(\lambda)(m_\sigma^\dr(\lambda) v^\eff_\sigma(\lambda))^2+\rho_\M(\lambda)(m_\M^\dr(\lambda) v^\eff_\M(\lambda))^2+\rho_\R(\lambda)\vartheta_\R(\lambda)(m_\R^\dr(\lambda)v^\eff_\R(\lambda))^2\, .
\ee
Similarly to the easy-axis case, the singularity of the bare energy placed at $(\sigma,\lambda)=(1,0)$ manifest as a singularity of the Drude weight at zero temperature $\sim -\log\beta$: the derivation of the divergence follow the same analysis of the easy-axis, and it is thus omitted.
In Fig. \ref{fig_S2}, we benchmark our TBA equations on the average magnetization and energy, with good agreement.

\bigskip
\textbf{The infinite-temperature limit.---} As in the axial regime, also in the planar case the filling function becomes rapidity independent in the infinite temperature limit $\beta\to 0$. The infinite-temperature TBA equations are easily written, but they are not easy to solve. We managed to find a solution by taking the semiclassical limit of the known solution for the quantum chain \eqref{eq_S_infQEP} and check a posteriori its correctness, obtaining
\be
\vartheta_\sigma=\frac{\mu^2}{\sinh^2(\mu\sigma)}\, ,\hspace{1pc}\vartheta_\R=e^{-\ell \mu}\frac{\sinh(\ell\mu)}{\mu}\, ,\hspace{1pc}\vartheta_\M= e^{-\mu \ell} \frac{\mu}{\sinh(\ell\mu)}\,.
\ee
To compute the total magnetization we follow another route compared with the easy-axis regime passing through the free-energy: the discussion is postpone to the next paragraph.

\bigskip
\textbf{The classical entropy.---} To compute the classical entropy, we pursue the same strategy we used in the easy-axis regime: we postulate the entropy has the form $\mathcal{S}=L\int \dd\lambda \left[\int \dd\sigma \rho_\sigma^t s_\sigma(\vartheta_\sigma)+\rho_\R^t s_\R(\vartheta_\R)+s_\M(\vartheta_\M)\right]+L C$ with $C$ a constant, and impose that the TBA equations \eqref{eq_TBA_P1}\eqref{eq_TBA_P2}\eqref{eq_TBA_P3} are recovered by minimizing the free energy.

This simple analysis gives
\be
s_\sigma(\vartheta_\sigma)=\vartheta_\sigma-1/\sigma^2-\log(\sigma^2\vartheta_\sigma)\vartheta_\sigma\, ,\hspace{2pc}s_\R(\vartheta_\R)=1-\log(\ell/\vartheta_\R)\, ,\hspace{2pc}s_\M(\vartheta_\M)=-\vartheta_\M\log \vartheta_\M+\vartheta_\M(1-\log\ell)\, .
\ee
Analogously to the easy-axis case, the constant $C$ is fixed by looking at the free energy in the infinite temperature limit

\be
F=L\int_0^{\pi/\rhop}\frac{\dd\sigma}{\sigma^2} \left(\min(\sigma,1)-\sigma\frac{\rhop}{\pi}\right)\left(1-\frac{\mu^2\sigma^2}{\sinh^2(\mu\sigma)}\right)+L\frac{\rhop}{\pi}\log\left(\frac{\mu \ell e^{\ell \mu}}{\sinh(\ell\mu)}\right)-L C=L(\mu+\log(\mu/\sinh\mu)-C)\, .
\ee
Hence, by consistency with the microscopic calculation, we ask $C=0$.
From the free energy, one can immediately derive the magnetization by noticing that $\langle M^3-M^3_\text{vac}\rangle=-\partial_\mu F$

\bigskip
\textbf{The spin-wave limit.---} The analysis of the spin-wave limit follows similar steps to those of the easy-axis. In Eqs. \eqref{eq_TBA_P1}\eqref{eq_TBA_P2}\eqref{eq_TBA_P3} we consider $\mu$ to be large. In this limit, the filling function $\vartheta_\M$ is exponentially suppressed and can be entirely neglected.
For the effective energy $\varepsilon_\sigma$, we can use the same approximation we made in the easy-axis regime and assume only small strings $\sigma\ll 1$ matter. In this limit we can replace $T_{\sigma,\sigma'}(\lambda)\simeq 2\min(\sigma,\sigma')\delta(\lambda)$. Similarly $T_{\sigma,\R}(\lambda)=-2\sigma \ell \delta(\lambda)$, however, as one can check a posteriori, $\log\epsilon_\R$ logarithmically diverges and this is negligible compared with the bare energy contribution in Eq. \eqref{eq_TBA_P1}. Hence, the strong-polarization approximation of Eq. \eqref{eq_TBA_P1} is formally identical to Eq. \eqref{eq_A_lowT}, with the only difference of computing $e^{(1)}(\lambda)$ in the easy-plane regime. In the TBA equations for radiation \eqref{eq_TBA_P3}, at the leading order one can approximate the effective energy with the bare source term $\vartheta_\R(\lambda)=\frac{1}{\beta e_\R(\lambda)+\mu m_\R}$ and entirely neglect dressing effects.

We should now retrieve the thermodynamics of spin waves: these modes receive contributions both from the already existing radiative modes, and, similarly to the easy-axis regime, from the collective effect of the solitons, analogously to Eq. \eqref{eq_soliton_resumming}.
To this end, we define $\partial_\lambda p^{(1)}(\lambda)=\lim_{\sigma\to 0}\sigma^{-1}\partial_\lambda p_\sigma(\lambda)$ and $e^{(1)}(\lambda)=\lim_{\sigma\to 0}\sigma^{-1} e_\sigma(\lambda)$, and the two momentum branches
\begin{eqnarray}
k_+(\lambda)&=&\int^\lambda \partial_{\lambda'} p^{(1)}(\lambda')=-2\arctan[\coth\lambda \tan(\rhop/2) ]\, .\hspace{1pc}k_+\in [-\pi,-\rhop]\cup[\rhop,\pi]\, ,\\
k_-(\lambda)&=&\int^\lambda \partial_{\lambda'} p_\R(\lambda')=-2\arctan[\tanh\lambda \tan(\rhop/2) ]\, .\hspace{1pc}k_-\in [-\rhop,\rhop]\, .
\end{eqnarray}
Together, the two branches $k_+$ and $k_-$ cover the whole Brillouin zone. By inverting the above relation $\lambda_+(k)$ and $\lambda_-(k)$ one obtains the spin waves dispersion
\be
e_\text{SW}(k)=\begin{cases} e^{(1)}(\lambda_+(k))\hspace{1pc} & k\in [-\pi,-\rhop]\cup[\rhop,\pi]\\
e_\R(\lambda_-(k))\hspace{1pc} & k\in [-\rhop,\rhop]\end{cases}\hspace{1pc}\Rightarrow \hspace{1pc}e_\text{SW}(k)=\frac{4\rhop}{\sin\rhop}(\cos\rhop-\cos k)\, .
\ee
This result agrees with the spin-wave analysis of Section \ref{sec_SW}.

\subsection{The isotropic limit}
\label{subsec_isotropic}

The isotropic point can be obtained as a limiting case of either the easy-axis, or easy-plane case. For the sake of simplicity, we consider the easy-axis and send the anisotropy to zer $\rhoa\to 0$. As it is clear from the scattering kernel \eqref{eq_STaxis}, a non-trivial limit can be obtained by rescaling the rapidities $\lambda\to \rhoa \lambda$ before taking the limit. The straightforward procedure leads to the TBA equations and Drude weight formally identical to Eqs. \eqref{eq_STBA} and \eqref{eq_Drude}, but now the rapidity domain covers the whole real axis $\lambda\in[-\infty,\infty]$, the bare energy and momentum are $\partial_\lambda p_\sigma=\pi T_{\sigma,1}(\lambda)$ and $e_\sigma(\lambda)=2\partial_\lambda p_\sigma(\lambda)$, while the scattering kernel at the isotropic point is
\be
T_{\sigma,\sigma'}(\lambda)=\frac{1}{\pi}\log\left[\frac{\lambda^2+\left(\frac{\sigma+\sigma'}{2}\right)^2}{\lambda^2+\left(\frac{\sigma-\sigma'}{2}\right)^2}\right]\, .
\ee
We do not repeat the analysis of the infinite temperature and spin-wave limits, since they are analogue to the easy-axis case. 
 
\subsection{The continuum Landau-Lifschitz}
\label{subsec_continuum}

We finally focus on the continuum limit of the classical spin chain Eq. (1).
The continuum limit can be taken at the level of TBA: for this reason, we introduce a lattice spacing $a \to 0$. In the continuum limit of the lattice model, one needs to zoom on slow quasiparticles carrying large magnetization: in order to get non-trivial scattering kernels, the anisotropy should be rescaled accordingly. For the sake of convenience, we consider separately the easy-axis and the easy-plane regimes: the isotropic point is straightforwardly obtained as a limiting case of the easy-axis regime.

\bigskip
\textbf{The easy-axis regime.---}
We consider the following rescaling. We introduce a label ``c" for those quantities that describe the continuum model
\be
\rhoa=a\varrho_{\text{ea};\cc}\, ,\hspace{1pc}\sigma=\sigma_\cc/a\, ,\hspace{1pc} e_{\sigma}(\lambda)=ae_{\cc; \sigma_\cc}(\lambda)\, ,\hspace{1pc}T_{\sigma,\sigma'}(\lambda)=\frac{1}{a}T_{\cc;\sigma_\cc,\sigma_\cc'}(\lambda)
\ee
Likewise, the temperature and chemical potential needs to be properly rescaled as 
$\beta=a^{-1}\beta_\cc$ and $\mu=a\mu_\cc$. 
The rapidity is not rescaled, and it still has values in the compact domain $\lambda\in[-\pi/2,\pi/2]$.

The $a\to 0$ limit can be safely taken, leading to the same form of the TBA equations and Drude weight as in the lattice version, provided the scattering kernels and dispersion laws are replaced with the continuum counterparts 
\be
\partial_\lambda p_{\cc;\sigma_\cc}(\lambda)=\frac{2\sinh(\varrho_{\text{ea};\cc}\sigma_\cc)}{\cosh(\varrho_{\text{ea};\cc} \sigma_\cc)-\cos(2\lambda)}\, ,\hspace{1pc}T_{\cc; \sigma_\cc,\sigma_\cc'}(\lambda)=\frac{1}{\pi\varrho_{\text{ea};\cc}}\log\left(\frac{\cosh((\sigma_\cc+\sigma_\cc')\varrho_{\text{ea};\cc})-\cos(2\lambda)}{\cosh((\sigma_\cc-\sigma_\cc')\varrho_{\text{ea};\cc})-\cos(2\lambda)}\right)\, ,
\ee
while the magnetization eigenvalue retains the same form $m_{\cc;\sigma_\cc}=2\sigma_\cc$ and the energy $e_{\cc,\sigma_\cc}=2\varrho_{\text{ea; c}}\partial_\lambda p_{\cc;\sigma_\cc}(\lambda)$. Notice that, in the continuum limit, the modes with divergent bare energy are now at $\sigma_c=0$ and $\lambda=0$.
By taking the continuum limit directly at the level of the spin-wave analysis of the lattice model in Section \ref{sec_SW}, and comparing with the spin-wave analysis of the continuum Hamiltonian, one can identify $\Delta=\varrho_{\text{ea};\cc}^2$.

\bigskip
\textbf{The easy-plane regime.---} In the continuum limit, the easy-plane features the same excitations that are also present in the lattice and the rapidity still covers the whole real axis $\lambda\in (-\infty,\infty)$. As in the easy-axis regime, we introduce a lattice spacing $a$ and the following rescaling
\be
\rhop=a\varrho_{\text{ep};\cc}\, ,\hspace{1pc}\sigma=\sigma_\cc/a\, ,\hspace{1pc} e_{\sigma}(\lambda)=ae_{\cc; \sigma_\cc}(\lambda)\, ,\hspace{1pc}e_{\R}(\lambda)=ae_{\cc; \R}(\lambda)
\ee
\be\label{eq_ST_cEP}
T_{\sigma,\sigma'}(\lambda)=\frac{1}{a}T_{\cc;\sigma_\cc,\sigma_\cc'}(\lambda)\, ,\hspace{1pc}T_{\sigma,\R}(\lambda)=\frac{1}{a}T_{\cc;\sigma_\cc,\R}(\lambda)\, ,\hspace{1pc}T_{\R,\R}(\lambda)=aT_{\cc;\R,\R}(\lambda)\, ,\hspace{1pc}T_{\M,\R}(\lambda)=T_{\cc;\M,\R}(\lambda)
\ee
It is also useful to introduce $\ell_\cc=\pi/\varrho_{\text{ep};\cc}$: notice that, due to the rescaling, $\sigma_\cc\in [0,\ell_\cc]$.
Taking the continuum limit $a\to 0$ one formally obtains the same TBA equations of the planar regime on the lattice, provided the continuum dispersion law and scattering kernels are used
\be
\partial_\lambda p_{\cc;\sigma_\cc}(\lambda)=\frac{2\sin(\varrho_{\text{ep};\cc}\sigma_\cc)}{\cosh(2\lambda)-\cos(\varrho_{\text{ep};\cc}\sigma_\cc)}\, ,\hspace{1pc}\partial_\lambda p_{\cc;\R}(\lambda)=-\frac{2\varrho_{\text{ep};\cc}}{\cosh(2\lambda)+\cos(\varrho_{\text{ep};\cc}\sigma_\cc)}
\ee
and $\partial_\lambda p_{\cc;\M}(\lambda)=0$, the energies are $e_{\cc,\sigma_\cc}(\lambda)=2\varrho_{\text{ep;} \cc}\partial_\lambda p_{\cc;\sigma_\cc}(\lambda)$, $e_{\cc,\R}(\lambda)=2\varrho_{\text{ea; }\cc}\partial_\lambda p_{\cc;\R}(\lambda)$ and $e_{\cc\text{;}\M}(\lambda)=0$ . The magnetization eigenvalue reads $m_{\cc;\sigma_\cc}=2\sigma_\cc$, $m_{\cc; \R}=2$ and $m_{\cc;\M}=2\ell_\cc$. Finally, the kernels are read from Eq. \eqref{eq_ST_cEP}.
From the spin-wave limit, we can identify $\Delta=-\varrho_{\text{ep};\cc}^2$.

\section{Numerical methods}
\label{sec_numerics}
In this section we provide a short account for the numerical methods we use, more precisely we describe the numerical discretization of the TBA in Section \ref{subsec_numTBA} and a short description of the Monte Carlo simulations in Section \ref{subsec_numMicro}. Working Mathematica notebooks for the numerical solution of the TBA and computation of the Drude weight are provided on Zenodo \cite{Zenodo}.

\subsection{Solving the TBA equations}
\label{subsec_numTBA}

For the sake of simplicity, we describe the discretization in the easy-axis case: the easy-plane regime is analogous.
The numerical solution of the TBA equations needs to correctly account for two different singularities: the first happens at small $\sigma$ and arbitary rapidities $\lambda$, where the filling function diverges as $\sim 1/\sigma^2$, the second is placed around $(\sigma,\lambda)=(1,0)$, where the bare energy and momentum become singular.
The first problem can be solved with a convenient reparametrization of the filling functions and dressing equations, along the lines of previous works on the focusing nonlinear Schr\"{o}dinger \cite{Koch2022} and sine-Gordon \cite{Koch2023} models. To this end, we define $\bar{\vartheta}_\sigma(\lambda)=\sigma^2\vartheta_\sigma(\lambda)$ and a new dressing operation that we denote as ``bold dressing" $(...)^\bdr$ such that $\sigma^2( g_\sigma)^\bdr=g_\sigma^\dr$.
The new dressing operation can be recast as the following linear equation
\be\label{eq_S5}
\sigma g_\sigma^\bdr(\lambda)=\sigma^{-1}g_\sigma(\lambda)-\sigma^{-1}\int_0^\infty \dd\sigma'\int_{-\pi/2}^{\pi/2}\dd\lambda' T_{\sigma,\sigma'}(\lambda-\lambda')\bar{\vartheta}_{\sigma'}(\lambda')g^\bdr_{\sigma'}(\lambda')\, .
\ee
For those bare functions $g_\sigma(\lambda)$ that linearly vanish at small $\sigma$, i.e. $g_\sigma(\lambda)\propto \sigma$ (like it happens for the magnetization, energy, momentum and their derivatives), the new dressing is smooth (due to the fact that also $T_{\sigma,\sigma'}(\lambda)$ vanishes as $\propto \sigma$). As a great advantage, in the new parametrization $\bar{\vartheta}_{\sigma'}g^\bdr_{\sigma'}(\lambda')$ is smooth and Eq. \eqref{eq_S5} is easier to discretize. Notice also that, very conveniently, $\rho_\sigma(\lambda)=\frac{1}{2\pi}\vartheta_\sigma(\partial_\lambda p_\sigma)^\dr=\frac{1}{2\pi}\bar{\vartheta}_\sigma(\partial_\lambda p_\sigma)^\bdr$. For what concerns the determination of the filling functions on thermal states, we use the parametrization of the filling function in terms of the effective energy $\varepsilon_\sigma(\lambda)$ according to Eq. \eqref{eq_STBA}: the effective energy can be considered a smooth function in the discretization.
We now approximate the integral equations with matrix-vector equations by building a grid in the $(\sigma,\lambda)$ space. In particular, we consider an independent cartesian discretization $\{\lambda_j\}_{j=1}^{N_\lambda}\otimes \{\sigma_i\}_{i=1}^{N_\sigma}$ where each point represent the edge of a small integration interval. It is important that rapidities are discretized independently from $\sigma$ to capture the asymptotic behavior of the scattering kernel $\lim_{\sigma,\sigma' \to 0}T_{\sigma,\sigma'}(\lambda)\propto \delta(\lambda)$. The discretizations $\{\lambda_j\}_{j=1}^{N_\lambda}$ and $\{\sigma_i\}_{i=1}^{N_\sigma}$ are non-uniform in space to better capture the dangerous points. Hence, $\{\sigma_i\}_{i=1}^{N_\sigma}$ is denser around zero and one, while $\{\lambda_j\}_{j=1}^{N_\lambda}$ is denser around zero.
We then discretize the filling as
\be
\bar{\vartheta}_{\sigma}(\lambda)\to \bar{\vartheta}(i,j)=\bar{\vartheta}_{\frac{1}{2}(\sigma_i+\sigma_{i+1})}\left(\frac{\lambda_j+\lambda_{j+1}}{2}\right)
\ee
and similarly for the other functions, which are now seen as a vector in the $(i,j)$ space with dimension $(N_\lambda-1)\times(N_\sigma-1)$.
The dressing equation \eqref{eq_S5} (and similarly the TBA \eqref{eq_STBA}) becomes a matrix equation as
\be
g^\bdr(i,j)=\left(\frac{\sigma_i+\sigma_{i+1}}{2}\right)^{-1}g_{\frac{\sigma_i+\sigma_{i+1}}{2}}\left(\frac{\lambda_j+\lambda_{j+1}}{2}\right)-\left(\frac{\sigma_i+\sigma_{i+1}}{2}\right)^{-1}\sum_{(i',j')} T[(i,j),(i',j')]\bar{\vartheta}(i',j')g^\bdr(i',j')\, .
\ee
Ideally, we define the discretized kernel as
\be\label{eq_Tmat}
T[(i,j),(i',j')]=\int_{\sigma_{i'}}^{\sigma_{i'+1}}\dd\sigma\int_{\lambda_{j'}}^{\lambda_{j'+1}}\dd\lambda \, T_{\frac{\sigma_i+\sigma_{i+1}}{2},\sigma}\left(\frac{\lambda_j+\lambda_{j+1}}{2}-\lambda\right)\, ,
\ee
but we approximate the integral by correctly accounting for the singular behavior of the kernel. In the easy-axis regime, we define the singular part of the kernel
\be
T^\text{Sing}_{\sigma,\sigma'}(\lambda)=\frac{1}{\pi\rhoa}\cos(\lambda) \log\left[\frac{(\sigma+\sigma')^2\rhoa^2+4\sin^2(\lambda)}{(\sigma-\sigma')^2\rhoa^2+4\sin^2(\lambda)}\right]\, ,
\ee
while the non-singular part is defined as the reminder $T^\text{NSing}_{\sigma,\sigma'}(\lambda)=T_{\sigma,\sigma'}(\lambda)-T^\text{Sing}_{\sigma,\sigma'}(\lambda)$. The singular part of the kernel has the following important features: \emph{i)} it captures all the singularities of $T_{\sigma,\sigma'}(\lambda)$, thus leaving $T^\text{NSing}_{\sigma,\sigma'}(\lambda)$ a smooth function, \emph{ii)} it has the correct periodicity over $\lambda$ and the correct asymptotics for small $\sigma$, i.e. $\lim_{\sigma\to 0}T^\text{Sing}_{\sigma,\sigma'}(\lambda)=0$, \emph{iii)} and finally it can be easily analytically integrated over both strings and rapidities. Therefore, in the definition of the discretized kernels \eqref{eq_Tmat}, we perform exactly the integral over the singular part and approximate by the midpoint rule the integration over the non-singular term
\begin{multline}
T[(i,j),(i',j')]\simeq(\sigma_{i'+1}-\sigma_{i'})(\lambda_{j'+1}-\lambda_{j'}) T^{\text{NSing}}_{\frac{\sigma_i+\sigma_{i+1}}{2},\frac{\sigma_{i'}+\sigma_{i'+1}}{2}}\left(\frac{\lambda_j+\lambda_{j+1}}{2}-\frac{\lambda_{j'}+\lambda_{j'+1}}{2}\right)+\\
\int_{\sigma_i}^{\sigma_{i+1}}\dd\sigma\int_{\lambda_j}^{\lambda_{j+1}}\dd\lambda \, T^{\text{Sing}}_{\frac{\sigma_i+\sigma_{i+1}}{2},\sigma}\left(\frac{\lambda_j+\lambda_{j+1}}{2}-\lambda\right)\, .
\end{multline}
With this discretization, the TBA equations are convergent and correctly reproduce the known limiting case (high temperature and small magnetization), and give results in agreement with numerical simulations. So far, we left pending the problem of the singularity in the bare energy and momentum at $(\sigma,\lambda)=(1,0)$, which must be carefully handled when computing the Drude weight. As an example, we focus on the dressed energy derivative $(\partial_\lambda e_\sigma)^\bdr$: the same analysis must be done for the dressed momentum derivative $(\partial_\lambda p_\sigma)^\bdr$ and the effective energy $\varepsilon_\sigma$ which parametrizes the filling. We notice that the singularity in $(\partial_\lambda e_\sigma)^\bdr$ is due to the bare term. It is thus convenient to define $(\partial_\lambda e_\sigma)^{\bdr,\text{NSing}}=(\partial_\lambda e_\sigma)^\bdr-\sigma^{-2}\partial_\lambda e_\sigma-\sigma^{-1} \partial_\lambda e^{(1)}$, where we recall the definition $e^{(1)}(\lambda)=\lim_{\sigma\to 0}\sigma^{-1}e_\sigma(\lambda)$: with this definition, $(\partial_\lambda e_\sigma)^{\bdr,\text{NSing}}$ is smooth everywhere and goes to a constant for $\sigma\to 0$. Hence, from the numerical solution of the dressing equation, we compute the discretized version of  $(\partial_\lambda e_\sigma)^{\bdr,\text{NSing}}$ and interpolate it: in this way, $(\partial_\lambda e_\sigma)^\bdr$ is split in a singular part that we analytically control and a non-singular part that comes from the interpolation. This expression and the analogue for $(\partial_\lambda p_\sigma)$ and $\varepsilon_\sigma$ are then plug into the integral expression for the Drude weight. In this case, a change of variables from $(\sigma,\lambda)$ to $(\sigma,E)$ is convenient, where we define $E=e_\sigma(\lambda)$: as we discussed when extracting the large temperaturure asymptotics of $\mathcal{D}$, the singularity of the dressed velocity is now placed at $E\to \infty$. In this space, we perform the integral first in the $\sigma$ axis at fixed $E$, and sample the curve up to an energy cutoff $E_c$ which is chosen in such a way it correctly reproduces the asymptotic form Eq. \eqref{eq_asympD}. Then, the curve is interpolated an numerically integrated up to $E<E_c$, while the tail $E>E_c$ is integrated analytically. This method gives convergent resultswhich are realiabele even in the large temperature limit.
A similar procedure is used for the easy-plane regime as well. 
On Zenodo \cite{Zenodo}, we provide commented Mathematica notebooks where this numerical scheme is implemented in practice.

\subsection{Microscopic simulations}
\label{subsec_numMicro}

The microscopic simulations consist in two steps: \emph{i)} sampling initial configurations from thermal ensemble using a Metropolis-Hasting method, and \emph{ii)} evolving each initial condition with a deterministic dynamics, computing the evolution of the spin current. The Drude weight is then computed by averaging over initial configurations. In what follows we discuss each of these steps in details.

\subsubsection{Monte-Carlo sampling}
We consider equilibrium ensembles 
\begin{equation}
	\rho_L(\beta, \mu) = \mathcal{Z}_L^{-1} e^{-\beta H  + \mu M^3}, \label{rho_def}
\end{equation} 
where $M^3 = \sum_{\ell=1}^L S^3_\ell$ is the total magnetization and 
\begin{equation}
	\mathcal{Z}_L = \int \dd \Omega^{\times L}\, e^{-\beta H + \mu M^3} \label{Z_def}
\end{equation}
is the partition function with the uniform measure $\dd \Omega^{\times L} = \prod_{\ell=1}^L \dd \Omega_\ell$. The expectation value of an observable $\mathcal{O}_\ell(t)$ in the ensemble \eqref{rho_def} is given by
\begin{equation}
	\langle \mathcal{O}_\ell(t) \rangle = \int \dd \Omega^{\times L}\, \mathcal{O}_\ell(t) \rho_L(\beta, \mu). \label{avg_def}
\end{equation}
 The ensemble \eqref{rho_def} is efficiently sampled by an implementation of the Monte-Carlo method:
\begin{enumerate}
	\item Sample an initial spin configuration $\{\textbf{S}\} = \{\textbf{S}_\ell\}_{\ell=1}^L$ from the uniform measure $\dd \Omega^{\times L}$.
	\item Compute the energy $H(\{\textbf{S}\})$ and magnetization $M^3(\{\textbf{S}\})$ of the spin configuration.
	\item Pick a random lattice site $1 \leq j \leq L$ and sample a randon direction $\textbf{N}$ from the uniform measure $\dd \Omega$.
 	\item Generate an alternative spin $\textbf{S}'_j = \textbf{S}_j \cos \theta + (\textbf{N}  \times \textbf{S}_j) \sin \theta + \textbf{N} (\textbf{N} \cdot \textbf{S}_j)(1-\cos \theta)$ where $\theta \ll 1$.
	\item Construct an alternative spin configuration $\{\textbf{S}'\} = \{\textbf{S}_\ell\}_{\ell=1}^{j-1} \cup \{ \textbf{S}'_j\} \cup \{\textbf{S}_\ell\}_{\ell=j+1}^L$.
	\item Compute the changes of energy $\delta H = H(\{\textbf{S}'\}) - H(\{\textbf{S}\})$ and magnetization $\delta M^3 = M^3(\{\textbf{S}'\}) - M^3(\{\textbf{S}\})$ of the alternative spin configuration relative to the unperturbed spin configuration.
	\item Update the spin configuration $\{\textbf{S}\} \leftarrow \{\textbf{S}'\}$ with probability $p = \min \{1, e^{-\beta \delta H + \mu \delta M^3} \}$.
	\item Repeat steps 3-7 until satisfactory convergence is obtained.
\end{enumerate}
Note that while the ensemble \eqref{rho_def} is an invariant measure for the continuous-time dynamics $\Phi$ \eqref{cont_map}, it is not an invariant measure for the discrete-time dynamics $\Phi_\tau$ \eqref{discrete_map} for $\tau > 0$ (see below). Our use of the continuous-time ensemble for the discrete-time dynamics makes the initial state effectively a non-equilibrium ensemble. However, we have verified that the resulting deviations from continuous-time dynamics at the used value of $\tau$ are within the estimated uncertainty interval.

\subsubsection{Time-evolution}
The evolution $\Phi: (\mathcal{S}^2)^{\times L} \to (\mathcal{S}^2)^{\times L}$ generated by the Hamiltonian (2) that maps the spins forwards in time
\begin{equation}
	\{\textbf{S}_\ell(t+t_0)\}_{\ell=1}^L = \Phi(t)\left[\{\textbf{S}_\ell(t_0)\}_{\ell=1}^L \right]. \label{cont_map}
\end{equation}
 is integrable, indeed it conserves an extensive tower of charges $\{Q_n\}_{n=1}^L$, the first one being the Hamiltonian $Q_1 \simeq H$. To avoid breaking its integrability by a direct time discretization, we instead use a discretization developed in \cite{Krajnik2021} that preserves integrability by construction. The discrete-time evolution map $\Phi_\tau: (\mathcal{S}^2)^{\times L} \to (\mathcal{S}^2)^{\times L}$
 \begin{equation}
 	\{\textbf{S}_\ell(T+T_0)\}_{\ell=1}^L = \Phi_\tau(T)\left[\{\textbf{S}_\ell(T_0)\}_{\ell=1}^L \right], \label{discrete_map}
 \end{equation}
depends on a time-step parameter $\tau \in \mathbb{R}$. Hamiltonian dynamics are  recovered in the continuous-time limit
\begin{equation}
	\lim_{\stackrel{\tau \to 0,\, T \to \infty}{T \tau = t}} \Phi_\tau(T) = \Phi(t).
\end{equation}
The discrete-time map $\Phi_\tau$ conserves an extensive tower of $\tau$-deformed charges $\{Q_n^{(\tau)} \}_{n=1}^L$ that reduce to their continuum-time counterparts, $\lim_{\tau \to 0}\{Q_n^{(\tau)} \}_{n=1}^L = \{Q_n \}_{n=1}^L$.
All reported simulations use the value $\tau= 0.03$ and are terminated before periodicity of the finite size of the system is manifest, $T_{\rm max} < L/2$. While the estimated values of dynamical quantities depend on the time-step parameter $\tau$, we have verified that results for smaller values of $\tau$ are contained within the estimated uncertainty interval.

\subsubsection{Expectation values}
We estimate expectation values \eqref{avg_def} by averaging over $N$ samples drawn from the ensemble \eqref{rho_def}
\begin{equation}
	\overline{\mathcal{O}_\ell(t)} = N^{-1} \sum_{n=1}^{N} \mathcal{O}^{[n]}_\ell(t), \label{O_est}
\end{equation}
where $\bullet^{[n]}$ denotes evaluation in the $n$-th sample. Uncertainty of the estimate \eqref{O_est} is extracted from the standard deviation of the samples
\begin{equation}
	\sigma^2_{\mathcal{O}_\ell(t)} = N^{-1} \sum_{n=1}^{N} \left[\mathcal{O}^{[n]}_\ell(t) - \langle \mathcal{O}_\ell(t) \rangle\right]^2. \label{O_std}
\end{equation} 
Since samples in Eq. \eqref{O_est} are independent, it follows by the central limit theorem that as the number of samples $N$ grows, the distribution of the average $\overline{\mathcal{O}_\ell(t)}$ approaches a Gaussian distribution centered at $\langle \mathcal{O}_\ell(t) \rangle$ with standard deviation
\begin{equation}
	\sigma_{\overline{\mathcal{O}}} = N^{-1/2}\sigma_{\mathcal{O}},
\end{equation} 
which gives the uncertainty interval (with standard score $z=1$) of the estimated average value
\begin{equation}
	\langle \mathcal{O}_\ell(t) \rangle \approx \overline{\mathcal{O}_\ell(t)} \pm \sigma_{\overline{\mathcal{O}}}.
\end{equation}

We extract the spin Drude weight using the Green-Kubo identity. We work in the continuous-time setting, the adaptation to the discrete-time setting is straightforward, see Ref. \cite{Krajnik2021} for details.
 Spin satisfies a continuity equation
\begin{equation}
	\partial_t S^3_\ell(t) + j_{\ell+1}(t)  - j_{\ell}(t) = 0,  
\end{equation}
where $j_\ell$ is the spin current density.  We define the total spin current as
\begin{equation}
	J(t) = \sum_{\ell=1}^L j_\ell(t).
\end{equation}
The finite-time finite-size Drude weight is given by the integral of the total spin current auto-correlation function
\begin{equation}
	\mathcal{D}_L(t) = t^{-1}L^{-1} \int_0^t \dd t\, \langle J(t)J(0) \rangle^c, \label{Drude_finite}
\end{equation}
where $\langle XY \rangle^c = \langle XY \rangle- \langle X \rangle\langle Y \rangle$ denotes the connected part of the correlation function. The Drude weight then correspond to the thermodynamic value of the long-time limit of \eqref{Drude_finite}
\begin{equation}
	\mathcal{D} = \lim_{t \to \infty}\lim_{L \to \infty} \mathcal{D}_L(t).
\end{equation}

\end{document}